\documentclass[sigconf, nonacm]{acmart}

\makeatletter
\def\@affiliationfont{\fontsize{9.5}{11}\selectfont\normalfont}
\makeatother

\newcommand\vldbdoi{XX.XX/XXX.XX}
\newcommand\vldbpages{XXX-XXX}
\newcommand\vldbvolume{19}
\newcommand\vldbissue{12}
\newcommand\vldbyear{2026}
\newcommand\vldbauthors{\authors}
\newcommand\vldbtitle{\shorttitle} 
\newcommand\vldbavailabilityurl{}
\newcommand\vldbpagestyle{empty} 

\usepackage{float}
\usepackage{graphicx}
\usepackage{enumerate}
\usepackage{enumitem}
\usepackage{subfigure}
\usepackage{multirow}
\usepackage{xcolor}
\usepackage{color}
\usepackage{mathtools}
\usepackage{epstopdf}
\usepackage{balance}
\usepackage{tabularx}
\usepackage{soul}

\usepackage[linesnumbered,vlined,ruled,noend]{algorithm2e}

\let\oldtcp\tcp
\renewcommand{\tcp}[1]{\oldtcp{\footnotesize\textcolor{blue}{#1}}}

\newcommand{\xy}[1]{\textcolor{cyan}{Xiaoying: #1}}
\newcommand{\todo}[1]{\textcolor{red}{[TODO] #1}}
\newcommand{\ww}[1]{\textcolor{blue}{Wentao: #1}}

\newcommand{\cut}[1]{}

\newcommand{\revision}[1]{\textcolor{black}{#1}}

\newtheoremstyle{tightfinding}
  {0.25em}   
  {0.25em}   
  {\itshape}         
  {}         
  {\bfseries}         
  {.}        
  {0.5em}         
  {}
\theoremstyle{tightfinding}

\newtheorem{observation}{Key Finding}

\newcommand{\stitle}[1]{\vspace{2pt}\textbf{#1}}

\begin{document}
\title{Evaluating the Practical Effectiveness of LLM-Driven Index Tuning on Microsoft SQL Server}


\settopmatter{authorsperrow=4}
\author{Xiaoying Wang}
\email{wangxiaoying@microsoft.com}
\affiliation{%
  \institution{Microsoft Research}
  \city{Redmond}
  \country{USA}
}

\author{Wentao Wu}
\email{wentao.wu@microsoft.com}
\affiliation{%
  \institution{Microsoft Research}
  \city{Redmond}
  \country{USA}
}

\author{Vivek Narasayya}
\email{viveknar@microsoft.com}
\affiliation{%
  \institution{Microsoft Research}
  \city{Redmond}
  \country{USA}
}

\author{Surajit Chaudhuri}
\email{surajitc@microsoft.com}
\affiliation{%
  \institution{Microsoft Research}
  \city{Redmond}
  \country{USA}
}



\begin{abstract}
Indexes are crucial for database performance. Index tuning, i.e., selecting appropriate indexes for a database workload, is an important problem. The state-of-the-art index tuning tools in the industry, e.g.,  Database Tuning Advisor (DTA) developed for Microsoft SQL Server, rely on a ``what-if'' API, which can estimate the cost of a query for a given index configuration. They take as input a SQL workload and constraints such as a storage bound, and search over the large space of index configurations to find one with low optimizer-estimated cost for the input workload. Large language models (LLMs) offer a different approach to index tuning, using knowledge they have learned from publicly available training data. However, the effectiveness of LLM-driven index tuning in comparison to today's index advisors, particularly on enterprise workloads, remains unclear.

In this paper, we study the practical effectiveness of LLM-driven index tuning on Microsoft SQL Server using both industrial benchmarks and real-world enterprise customer workloads, and compare it with DTA. Our results show that while LLMs in several cases identify configurations that significantly outperform those found by DTA in terms of execution time, they suffer from high variance in index recommendation quality. Furthermore, index recommendations from the LLM are often substantially worse than DTA in terms of optimizer-estimated cost, making it challenging to extend cost-based index advisors such as DTA to leverage LLMs for index tuning. We point to some areas of future work that may be important for robustly leveraging LLMs for index tuning.

\end{abstract}

\sloppy
\maketitle

\pagestyle{\vldbpagestyle}
\begingroup\small\noindent\raggedright\textbf{PVLDB Reference Format:}\\
\vldbauthors. \vldbtitle. PVLDB, \vldbvolume(\vldbissue): \vldbpages, \vldbyear.\\
\href{https://doi.org/\vldbdoi}{doi:\vldbdoi}
\endgroup
\begingroup
\renewcommand\thefootnote{}\footnote{\noindent
This work is licensed under the Creative Commons BY-NC-ND 4.0 International License. Visit \url{https://creativecommons.org/licenses/by-nc-nd/4.0/} to view a copy of this license. For any use beyond those covered by this license, obtain permission by emailing \href{mailto:info@vldb.org}{info@vldb.org}. Copyright is held by the owner/author(s). Publication rights licensed to the VLDB Endowment. \\
\raggedright Proceedings of the VLDB Endowment, Vol. \vldbvolume, No. \vldbissue\ %
ISSN 2150-8097. \\
\href{https://doi.org/\vldbdoi}{doi:\vldbdoi} \\
}\addtocounter{footnote}{-1}\endgroup

\ifdefempty{\vldbavailabilityurl}{}{
\vspace{.3cm}
\begingroup\small\noindent\raggedright\textbf{PVLDB Artifact Availability:}\\
The source code, data, and/or other artifacts have been made available at \url{\vldbavailabilityurl}.
\endgroup
}

\section{Introduction}\label{sec:intro}

Index tuning, the problem of deciding which indexes to create for a given database workload, is an important and challenging problem that has been studied extensively~\cite{Whang85,ChaudhuriN97,dta,ValentinZZLS00,BrunoC05,DashPA11,SchlosserK019,KossmannHJS20,WuWSWNCB22,Wii,Esc,SiddiquiWNC22,ml-index-tuning-overview}.
Today's index tuning tools, aka \emph{index advisors} deployed by commercial and open-source database systems such as Microsoft SQL Server~\cite{ChaudhuriN97,AgrawalCKMNS04}, IBM DB2~\cite{ValentinZZLS00}, and PostgreSQL~\cite{dexter-2}, adopt a cost-based architecture where the goal is to minimize the cost of the input SQL workload \emph{estimated} by the database query optimizer using a ``what-if'' API. 
Although these cost-based index advisors have been proven effective in practice, inaccuracies in cost estimates can lead to suboptimal index recommendations~\cite{WuCZTHN13,DingDM0CN19}, sometimes even resulting in  query performance regressions (QPRs)~\cite{DasGIJJNRSXC19,WuDXNC25}. 
To mitigate the impact of inaccurate cost estimates on index tuning, previous work has explored techniques such as QPR detection or improvements to underlying cost models~\cite{DingDM0CN19,WuDXNC25,DBLP:journals/pvldb/MarcusP19,DBLP:conf/sigmod/MaDDS20}. 
However, the effectiveness of these remedies remains limited due to fundamental challenges in cost estimation~\cite{LeisGMBK015,IoannidisC91,Lohman-critique}.

Large language models (LLMs) offer an alternative approach to index tuning. 
LLMs have shown promise in other database performance tuning tasks, such as query rewriting~\cite{DBLP:conf/cidr/NarasayyaC26,ma-etal-2023-query,ye-etal-2023-enhancing,dharwada2025query,liu2024genrewrite,llmr2} and query optimization~\cite{abs-2411-02862,DBLP:journals/pacmmod/TanZLYPCMZR25,yao2025query}, suggesting that it can capture and apply useful domain knowledge without explicit cost models. 
Several recent studies have taken initial steps toward the application of LLMs to index tuning~\cite{GiannakourisT25,LLMIdxAdvis,MAAdvisor}, exploring different task formulations and prompt designs, and reporting promising early results. 
However, these studies primarily evaluated LLMs on open benchmarks that are likely included in LLM's training data. 
As a result, the effectiveness of LLMs when applied to real-world enterprise workloads is not well understood, 
particularly those involving complex queries with views, common table expressions (CTEs), nested sub-queries, etc. or databases with a large number of manually created indexes --- characteristics that open benchmarks typically lack. 
Moreover, the benefits of LLMs compared to a state-of-the-art commercial index tuner, such as Microsoft SQL Server Database Tuning Advisor (DTA)~\cite{dta}, remain unclear, as the evaluations in previous work were limited to PostgreSQL and comparison with simplified baselines.

In this paper, we study LLM-driven index tuning using both public benchmarks and \emph{real-world enterprise customer workloads}. We evaluate its effectiveness in comparison with DTA on Microsoft SQL Server.
We focus on pre-trained LLMs in an end-to-end setting, given only basic information about the database and workload. 
Unlike previous work that relies mainly on estimated cost from the query optimizer~\cite{KossmannHJS20,10.14778/3675034.3675035}, our evaluation is based on the actual \emph{execution time} of each query. 
Our study covers the following aspects:
\begin{itemize}[leftmargin=*]
    \item \emph{Single-query vs. Multi-query Workloads.} We evaluate index tuning for both single-query and multi-query workloads, both of which are important scenarios in practice. 
    \item \emph{Constrained vs. Non-constrained Tuning.} Constraints (e.g., the maximum number of indexes or storage space allowed) play an essential role in index tuning as they can change the search space of the index tuner. Both constrained and non-constrained index tuning have applications in practice~\cite{ChaudhuriN97,ValentinZZLS00,KossmannHJS20,WuWSWNCB22}. 
    \item \emph{Synthetic Benchmarks vs. Real-world Workloads.} Existing evaluations are largely based on public benchmarks, which are available for LLM training. Real-world enterprise customer workloads, on which LLMs have not been trained, therefore provide a new test to assess the effectiveness of LLM-driven index tuning. 
\end{itemize}

\vspace{-0.5em}

In our experiments, we invoke the LLM five times for each query or workload. For each invocation, we implement the LLM's recommendation and execute the query/workload. Due to non-determinism inherent in LLMs, we observe that even for the same query, LLM's recommendations change significantly across invocations. Note that we invoke DTA only once since it is deterministic. The key findings from our experimental study are:

\vspace{0.5em}

\stitle{LLM-recommended indexes provide complementary benefits compared to DTA.} 
When considering the \emph{best} recommendation across the five invocations, for a significant fraction of single-query workloads (approximately 67\%), we find that the LLM can identify a configuration that is comparable to that returned by DTA (Section~\ref{sec:single-query-overview}). These results hold for both industry benchmarks and real-world workloads.
We also find that the LLM often proposes fewer indexes than DTA (Section~\ref{sec:single-query-index-usage}). 
Moreover, for 31\% of queries, the LLM identifies configurations that substantially outperform those recommended by DTA ($\geq20$\% faster). Such queries are primarily those where DTA's recommendation is misled by inaccurate optimizer cost estimates (Section~\ref{sec:single-query-estcost}). 
For multi-query workloads, DTA delivers more consistent performance improvements overall compared to the LLM. However, the LLM still suggests configurations that significantly outperform DTA in two of the five workloads evaluated in our experiments (Section~\ref{sec:multi-query-workloads:overview}). 
 
\vspace{0.5em}
\stitle{LLM exhibits high performance variance, with worst-case outcomes significantly worse than DTA.} 
LLM-driven index tuning exhibits substantial performance variance across invocations for the same query, and its worst-case outcomes can significantly underperform DTA. In some cases, these recommendations can also lead to severe QPRs (Sections~\ref{sec:single-query-workloads:robustness} and~\ref{sec:multi-query-workloads:overview}). 
This variance occurs across workloads as well. For example, in some workloads, even the \emph{best} index configuration found by the LLM is significantly worse than DTA. We also observed that poor LLM recommendations became more frequent as workload size increases (Section~\ref{sec:multi-query-workloads:real-m}).
These results suggest that, despite their promise, it is not prudent to directly adopt LLM-generated recommendations in practice. The next two findings are motivated by the goal of evaluating if LLMs can be \emph{robustly} used in index tuning, i.e., without incurring the high risk of performance degradation. 


\vspace{0.5em}
\stitle{Directly integrating LLM recommendations into DTA often leads to performance degradation.} 
We evaluate the feasibility of expanding DTA’s pool of candidate indexes with LLM-recommended indexes, while leaving the rest of DTA unchanged. In principle, this approach expands DTA’s search space and may allow it to identify configurations with lower estimated costs while retaining the deterministic behavior of DTA. 
However, for single-query workloads, our results show that this strategy is ineffective in practice. The final configurations selected by DTA from the augmented pool of candidates is often unchanged, or leads to performance degradation when implemented, again due to inaccuracies in the optimizer's cost estimation (Section~\ref{sec:lesson:integration}).
For multi-query workloads, the LLM tends to directly identifies indexes that benefit multiple queries. This is in contrast to DTA, which first optimizes each query independently to identify candidates and then merges candidates from different queries to include new candidates that are promising at the workload level. 
Indexes identified by the LLM are not always considered by DTA and, in two out of the five workloads, expanding DTA's pool of candidate indexes with LLM-recommended indexes improves DTA's quality significantly (Section~\ref{sec:multi-query-workloads:real-d} and~\ref{sec:lesson:integration}). 

\vspace{0.5em}
\stitle{Effective heuristics can be distilled from LLM.} 
When analyzing LLMs' recommendations and their reasoning process on single-query workloads, we observed a few recurring patterns in their choice of indexes (Section~\ref{sec:llm-efficacy:reasoning}). 
Based on this observation, we develop a simple rule-based index tuner that aims to distill these patterns into a deterministic program (Section~\ref{sec:llm-efficacy:simple}). 
Interestingly, we find that the heuristic index tuner provides most of the benefits of LLM-recommended indexes, particularly for queries where DTA underperforms LLM (Section~\ref{sec:llm-efficacy:simple_eval}). Although more experiments are required to evaluate whether this approach is effective across different prompts and models, this experiment serves as a ``proof-of-concept'' that meaningful value can be derived from analyzing LLM's reasoning without necessarily incurring the high variance that accompanies LLMs.

\setlength{\tabcolsep}{3pt}
\begin{table}[t]
\caption{Summary of database and workload properties.} 
\vspace{-1em}
\label{tab:databases}
\scriptsize
\centering
\begin{tabular}{|l|r|r|r|r|r|r|r|r|}
\hline
\textbf{Name} & \textbf{DB Size} & \textbf{\#} \textbf{Queries} & \textbf{\#} \textbf{Tables} & \textbf{\#} \textbf{Joins} 
& \textbf{\#} \textbf{Scans}
& \textbf{\# Ops (Avg/Max)} \\
\hline
\hline
\textbf{TPC-H}	&	sf=10	&	22	&	8	&	2.8	&	3.7 &   19.1 / 38	\\
\textbf{Real-D}	&	587GB	&	29	&	7912	&	15.6	&	17 &   71.7 / 197	\\
\textbf{Real-M}	&	26GB	&	40	&	474	&	20.2	&	21.7 &   71.7 / 183	\\
\textbf{Real-R}	&	100GB	&	24	&	20	&	6.5	&	7.2   &   39.3 / 210	\\
\textbf{Real-S}	&	40GB	&	12	&	32	&	7.3	&	9.7    &   40.9 / 67	\\
\hline
\end{tabular}
\vspace{0.5em}
\end{table}

\vspace{0.5em}

In the following, we summarize the implications of our study for database practitioners and researchers and point to opportunities for future work.
\vspace{-0.5em}
\begin{itemize}[leftmargin=*]
    \item Although LLMs demonstrate potential for providing good-quality index recommendations on industry benchmarks and real-world enterprise workloads, due to the high variance in quality, \emph{safely} realizing their potential in practice remains an open problem.
    \item A straightforward integration of LLM-recommended indexes into cost-based index tuning advisors such as DTA is ineffective in practice. Exploring alternative ways of leveraging the best of both approaches, e.g., with prompts that may help limit the downside, or using index advisors as a tool and having LLM play the role of an orchestrator, is an interesting area of future work.
    \item \emph{Validation} of index recommendations has previously been used to prevent regressions by evaluating a configuration after deployment and workload execution~\cite{ChakkappenKMKSZZLZ25,WuDXNC25}. However, index creation and workload execution can be expensive and disruptive. Since LLMs can identify high-quality indexes but exhibit high variance, validation is essential for leveraging LLM-driven index tuning. Thus, developing more efficient and less disruptive validation mechanisms, such as those enabled by branching techniques~\cite{neon-db-branching}, is an important direction for future index tuning.
    \item Another direction is to harness the potential of LLM-driven index tuning while reducing its variance. One approach is to develop heuristic techniques by incorporating domain knowledge distilled from LLMs. It is also interesting to investigate whether models fine-tuned for index tuning can outperform pre-trained LLMs by achieving more consistent quality.
\end{itemize}

\section{Methodology}
\label{sec:methodology}

In this section, we present the methodology used in our study. 

\subsection{Benchmark and Customer Workloads}
Unlike existing work that has been primarily evaluated using public benchmarks with synthetic data and queries, we focus on real-world customer workloads in our study.
Table~\ref{tab:databases} summarizes the properties of the workloads that we used, including schema complexity (e.g., the number of tables), query complexity (e.g., the average and/or maximum number of joins, table scans, and operators per query), and the sizes of database and workload. 

We adopt four customer workloads with diverse properties. 
These workloads consist of real‑world analytical queries that make intensive use of features such as CTEs and views, and the underlying tables contain numerous pre-existing indexes that were manually created by human experts. 
We also include the standard \textbf{TPC-H} benchmark 
with a scaling factor (sf) of 10.


\vspace{-0.5em}
\subsection{Experimental Setups}

\begin{figure}[t]
\centering\vspace{-1em}
    \includegraphics[clip, trim=1.4cm 1cm 2.4cm 1.2cm, width=\columnwidth]{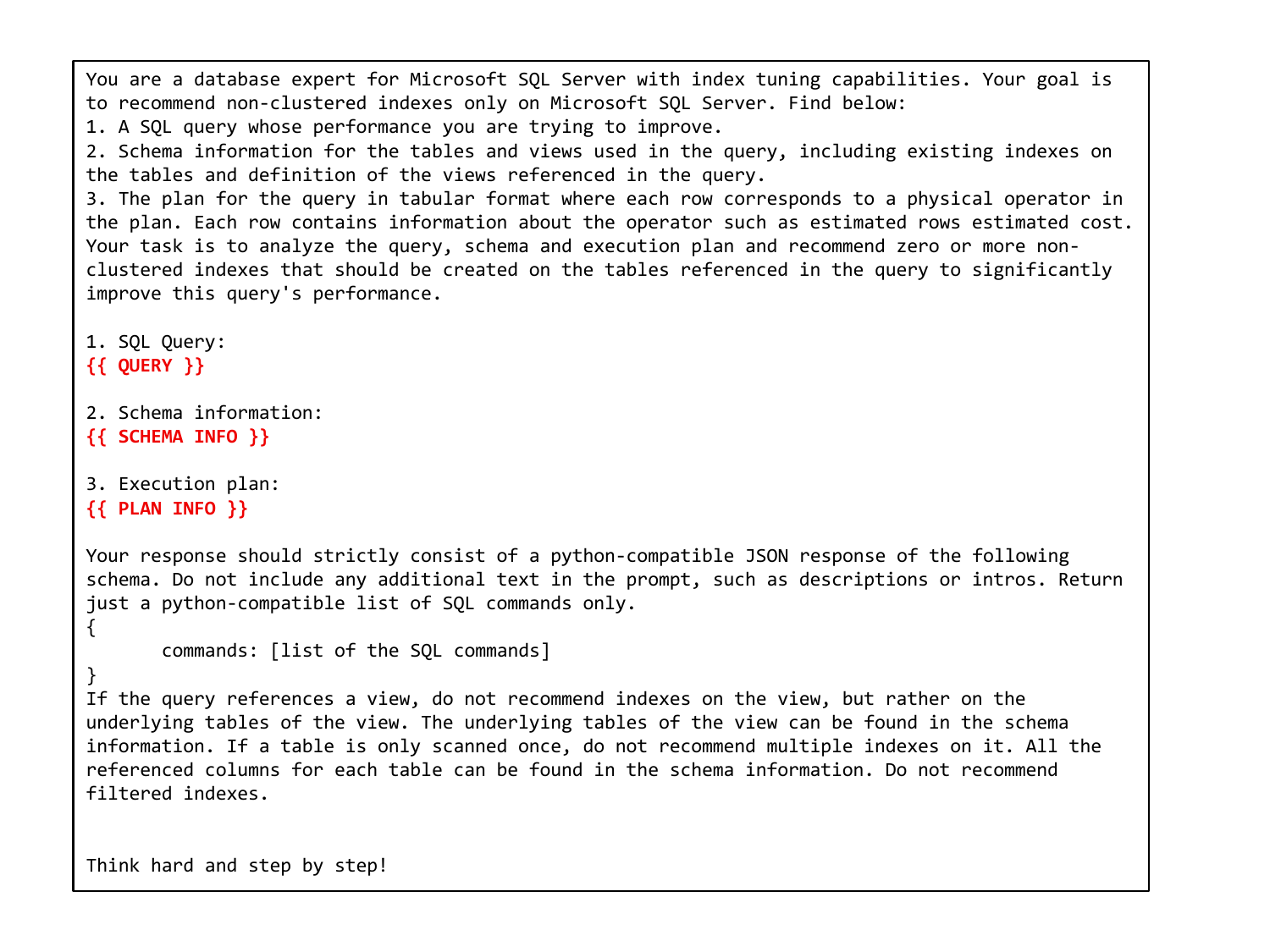}
\vspace{-2em}
\caption{Single-query workload prompt template.} 
\vspace{1em}
\label{fig:prompt-query}
\end{figure}

\subsubsection{Evaluation Baseline}
To evaluate the effectiveness of LLM-driven index tuning on real-world queries, we adopt Microsoft Database Tuning Advisor (DTA)~\cite{dta,AgrawalCKMNS04} as the baseline in this study. DTA is a commercial industrial-strength index tuner that represents the current state of the art (SOTA) in classic cost-based index tuning. Thus, it serves as a strong reference point for comparing LLM-driven index tuning.

DTA adopts a cost-based software architecture that contains three major components: (1) \emph{workload parsing and analysis}, which analyzes input workload queries to identify \emph{indexable columns}~\cite{ChaudhuriN97}, such as columns appearing in filter or join conditions; (2) \emph{candidate index generation}, which determines the key columns and included columns of potentially beneficial indexes as well as the orders of the columns; and (3) \emph{configuration enumeration}, which selects a subset (a.k.a. a configuration) from the candidate indexes that can minimize the estimated workload execution cost while adhere to the given index tuning constraints.
To estimate the cost of a query (and thus the workload) with an index configuration, 
DTA uses the ``what-if'' API~\cite{ChaudhuriN98}, which is an extended functionality of the query optimizer that can perform this cost estimation \emph{without} materializing the indexes to the underlying storage system. 


\revision{Recent studies have proposed various extensions to DTA such as reducing query performance regression~\cite{DingDM0CN19,DBLP:journals/tkde/Wu25} and improving its scalability when processing large workloads~\cite{SiddiquiJ00NC22,SiddiquiWNC22,WuWSWNCB22,Wii,Esc}. 
These techniques have not been used in the production DTA studied in this paper, and their integration with DTA remains future work.}

\subsubsection{Performance Metric}
Rather than using estimated cost~\cite{KossmannHJS20,10.14778/3675034.3675035}, we adopt execution time as our main evaluation metric. 
For each configuration, we execute each query in isolation five times, with each run capped at five minutes (300 seconds), and report the median execution time. 
For multi-query workloads, we use the total execution time of all queries in the workload.

\subsubsection{Hardware and Software Platform}
We use Microsoft SQL Server 2022 as the database system for evaluation. All experiments are conducted on a standard Azure M32bds v3 virtual machine~\cite{azurevm} with 32 vCPUs and 256 GB memory, running Windows Server 2022.

\begin{figure}[t]
\centering\vspace{-1em}
    \includegraphics[clip, trim=2.1cm 5.5cm 4.3cm 4.5cm, width=\columnwidth]{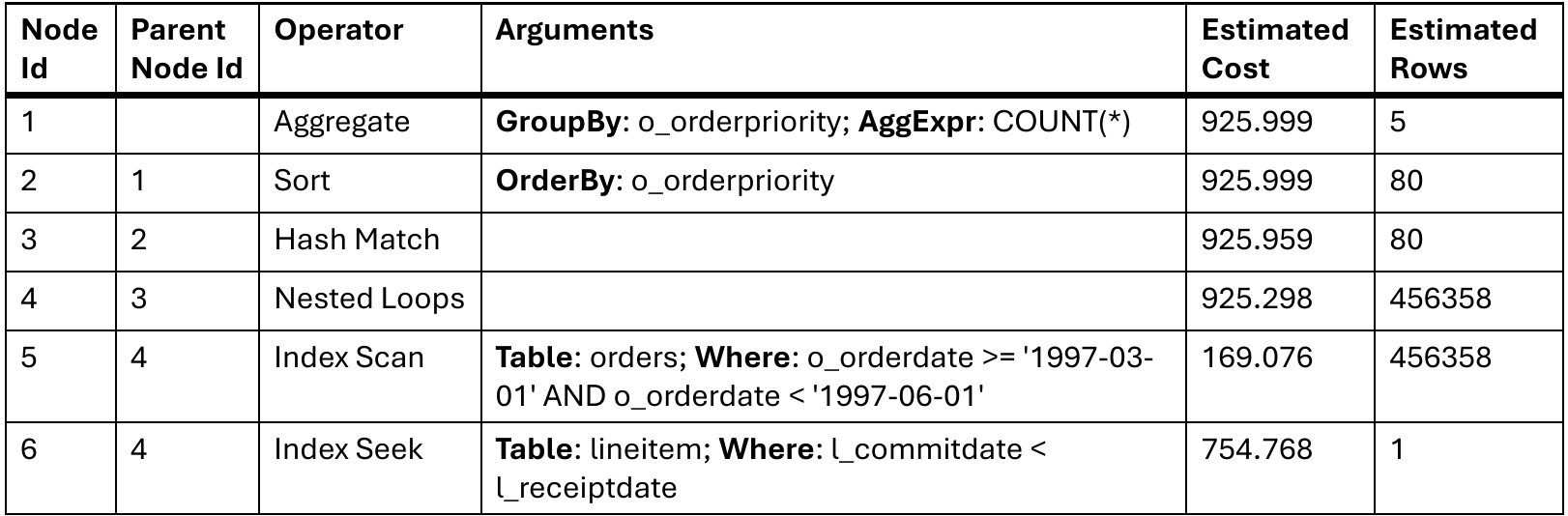}
\vspace{-2em}
\caption{Prompt plan example (TPC-H q04, simplified).} 
\label{fig:plan-example}
\end{figure}

\subsection{Setups of LLM-Driven Index Tuning}

We tried various LLMs, such as DeepSeek-R1~\cite{guo2025deepseek}, Qwen3~\cite{yang2025qwen3}, GPT-4o~\cite{hurst2024gpt}, and GPT-5~\cite{gpt-5}. We observe substantial variance in the effectiveness of different LLMs, with GPT-5 performing the best among all models evaluated in our experiments. 
Therefore, \emph{all evaluation results reported in this paper are based on GPT-5}. 
For each prompt, we invoke GPT-5 five times independently and collect the responses.
In the remainder of this paper, \emph{we use the two terms GPT-5 and LLM interchangeably}.

\subsubsection{Single-query Workload Tuning}

Figure~\ref{fig:prompt-query} presents the prompt template for tuning single-query workloads. 
It begins with instructions that describe the input information provided and the optimization task.
It ends with specifications of the output format and additional constraints, such as prohibiting index creation on views. 
Query-specific information that needs to be included in the prompt consists of the following: (1) the SQL text of the query, (2) the database schema, and (3) the current query execution plan based on the original index configuration.
This information is represented by the placeholders $\{\{\cdots\}\}$ in Figure~\ref{fig:prompt-query}.
\vspace{-0.5em}
\paragraph*{Database Schema}
The information about the database schema contains all \emph{base tables} that are referenced by the query, along with their cardinalities, included columns, and \emph{pre-existing indexes}. Definitions of all \emph{views} referenced by the query are also included. 

\vspace{-0.5em}
\paragraph*{Query Execution Plan}
Following previous work~\cite{DBLP:conf/cidr/NarasayyaC26}, the query plan with the original index configuration (i.e., pre-existing indexes) is represented in tabular form, where each row corresponds to a \emph{physical} operator and includes its detailed description (e.g., filter or join condition), estimated cost, and estimated cardinality. 
To ensure a fair comparison with DTA, query plans are obtained using the Microsoft SQL Server command ``\textsf{SET SHOWPLAN\_XML ON}'' without executions. 
As an example, Figure~\ref{fig:plan-example} shows the query plan representation for \textbf{TPC‑H} query 4. The plan uses a nested loop followed by a hash match to implement the semi-join, then sorts and aggregates the qualifying rows. Details such as parallelism are omitted in this example for the sake of clarity.

\vspace{-0.5em}
\paragraph*{Remark}
Our goal is not to identify an optimal prompt, but to provide LLM with basic information that a human expert would need to make informed index recommendations. 
In this study, we focus on evaluating the fundamental capability of LLM for index tuning, given only rudimentary information. 
More variations of LLM-driven index tuning are discussed in Section~\ref{sec:discussion}. 

\begin{figure}[t]
\centering
    \includegraphics[clip, trim=1.9cm 3.4cm 1.9cm 2.25cm, width=\columnwidth]{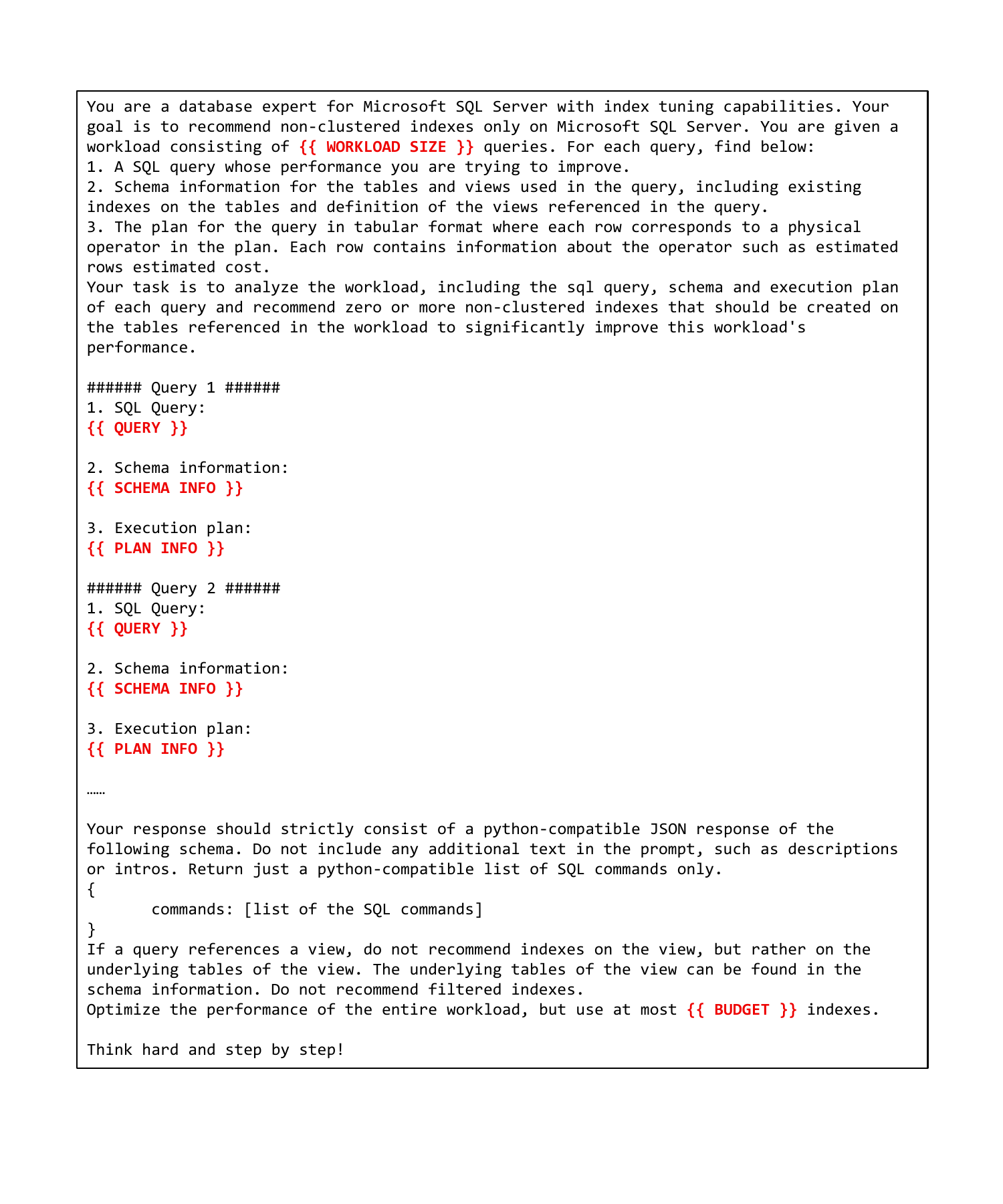}
\vspace{-1em}
\caption{Multi-query workload prompt template.} 
\vspace{1em}
\label{fig:prompt-workload}
\end{figure}

\subsubsection{Multi-query Workload Tuning}
\label{sec:methodology:multi-query}
For each query in the workload, we generate query-specific information using the same procedure as for single-query workloads and then concatenate these representations. 
Figure~\ref{fig:prompt-workload} presents the prompt template used for tuning multi-query workloads.
We make a few modifications to the instructions to accommodate the multi-query tuning scenario, incorporating additional information such as the workload size and the number of indexes allowed.

Unlike existing work~\cite{GiannakourisT25,LLMIdxAdvis} that compresses input workload information to stay within LLM's token limit, we deliberately retain the complete information for each query in the prompt, for two reasons. 
First, GPT-5 supports up to one million tokens, which is sufficient for many real-world customer workloads. 
Second, input compression may eliminate critical details of the workload, resulting in degraded performance and obscuring the true capabilities of LLM that we aim to evaluate.

\section{Single-query Workloads}
\label{sec:single-query-workloads}

In real-world applications, single-query workloads are common. 
For example, Microsoft SQL Server has a feature such that the query optimizer can automatically suggest ``missing indexes'' for a given query that would significantly improve the query performance if these indexes were created~\cite{sql-server-missing-indexes}.
Moreover, they are special cases of more general multi-query workloads that will be studied in Section~\ref{sec:multi-query-workloads}.
In fact, DTA uses a \emph{two-phase greedy search} algorithm~\cite{ChaudhuriN97,WuWSWNCB22} where it breaks the overall index selection problem into two search phases: (1) query-level search and (2) workload-level search.
The query-level search phase addresses exactly the single-query workload tuning problem.

We do not enforce constraints such as the maximum storage space allowed for LLM on single-query workloads.
By default, DTA sets the storage space constraint as 3$\times$ of the database size, which is more than sufficient for any single query and would not result in premature recommendations due to space constraints.

\vspace{-0.5em}
\subsection{Overview of Evaluation Results}\label{sec:single-query-overview}

\begin{figure*}
\centering\vspace{-1em}
\setcounter{subfigure}{0}
\subfigure[\textbf{TPC-H}]{ \label{fig:llm-vs-dta:tpch}
    \includegraphics[width=0.47\textwidth]{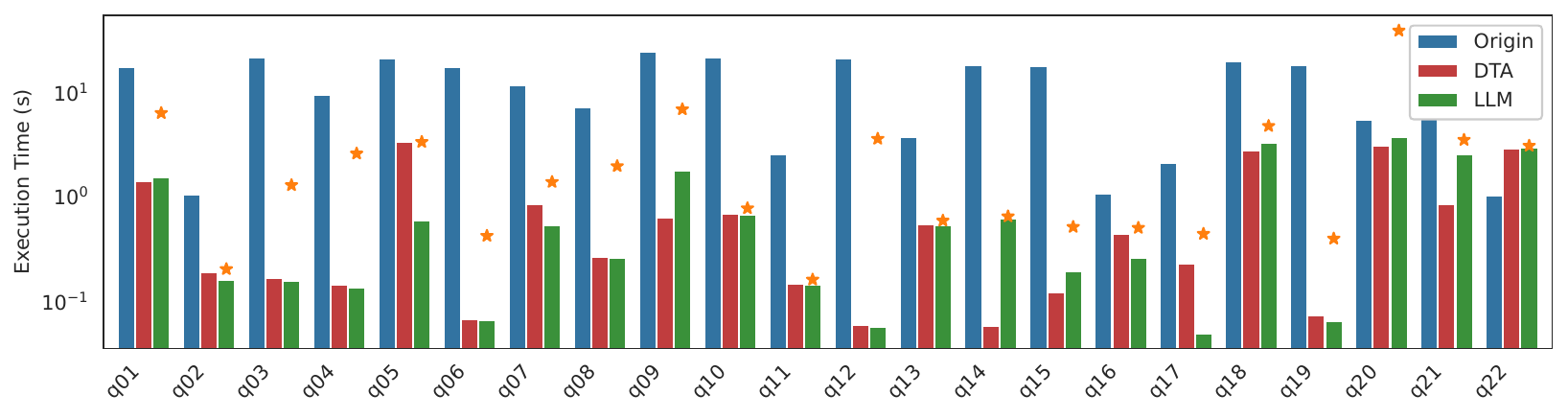}}
\setcounter{subfigure}{3}
\subfigure[\textbf{Real-R}]{ \label{fig:llm-vs-dta:real-r}
    \includegraphics[width=0.51\textwidth]{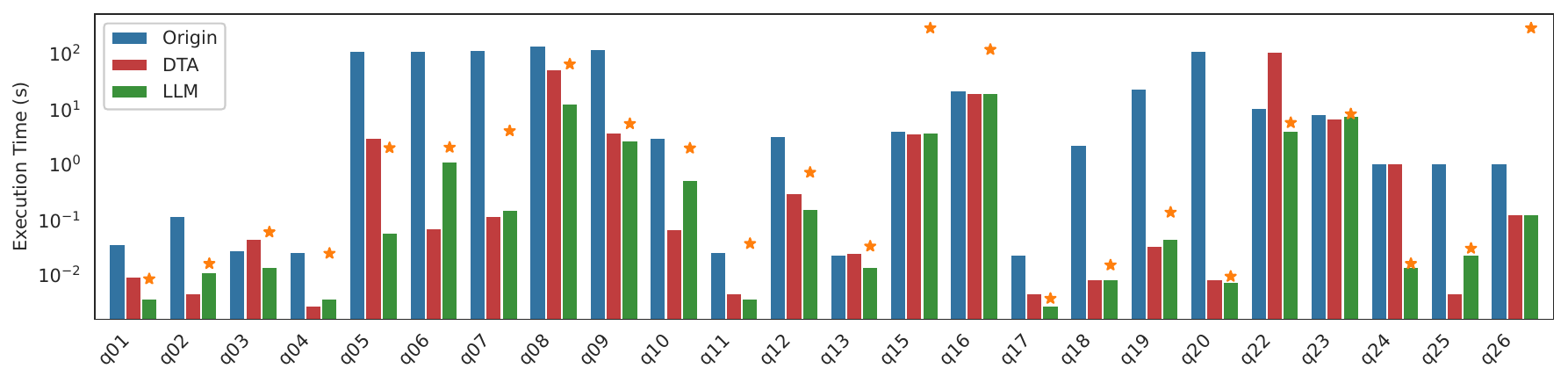}}
\setcounter{subfigure}{1}
\subfigure[\textbf{Real-D}]{ \label{fig:llm-vs-dta:real-d}
    \includegraphics[width=0.69\textwidth]{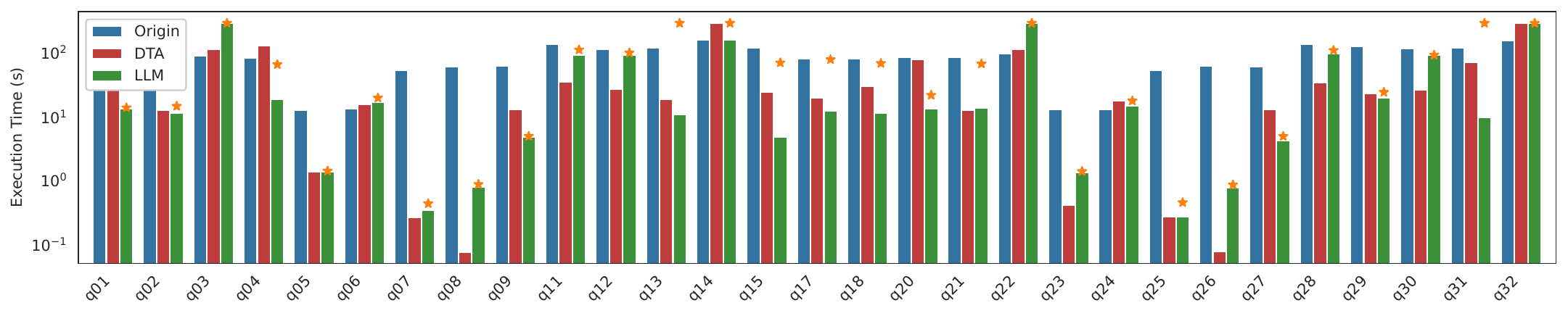}}
\setcounter{subfigure}{4}
\subfigure[\textbf{Real-S}]{ \label{fig:llm-vs-dta:real-s}
    \includegraphics[width=0.28\textwidth]{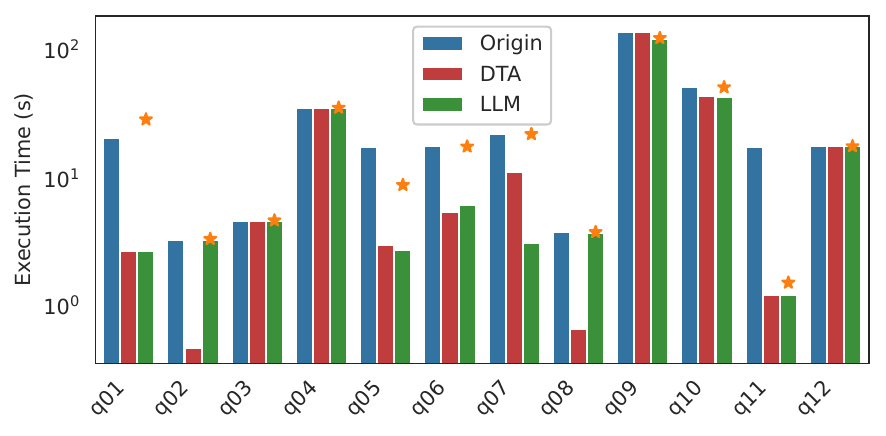}}
\setcounter{subfigure}{2}
\subfigure[\textbf{Real-M}]{ \label{fig:llm-vs-dta:real-m}
    \includegraphics[width=0.98\textwidth]{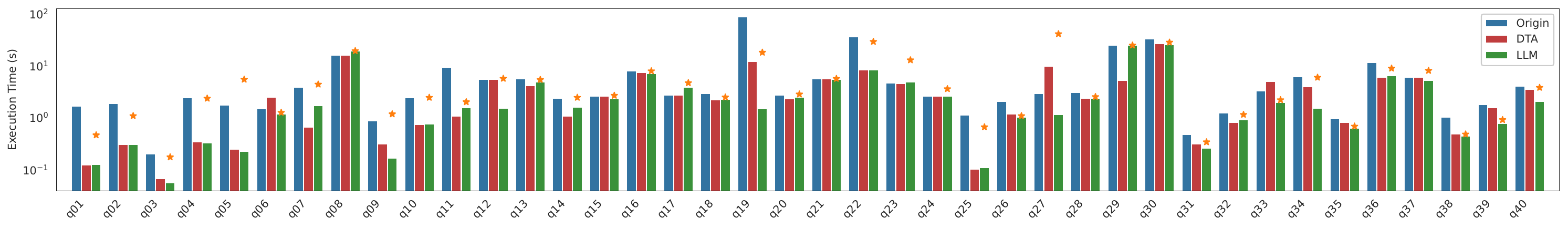}}
\vspace{-1.5em}
\caption{LLM-driven index tuning (best) vs. DTA for tuning single-query workloads (marker denotes the worst run of LLM).}
\label{fig:llm-vs-dta}
\vspace{-1.5em}
\end{figure*}

Figure~\ref{fig:llm-vs-dta} presents the evaluation results when comparing GPT-5 with DTA for tuning single-query workloads, in terms of query execution time with the recommended indexes. 
\revision{To illustrate the best potential of LLM, in the analysis from Section~\ref{sec:single-query-overview} to Section~\ref{sec:single-query-estcost}, we use the best-performing recommendation among the five LLM responses (denoted by the green bar in Figure~\ref{fig:llm-vs-dta}) as the outcome of LLM-driven index tuning. 
We also discuss worst-case results in Section~\ref{sec:single-query-workloads:robustness} but defer other results (e.g., from the first and median responses) to the full version of the paper~\cite{full-version}.}
The average response time of GPT‑5 is 1.1, 1.7, 1.9, 2.2, and 2.0 minutes on queries in \textbf{TPC-H}, \textbf{Real-M}, \textbf{Real-D}, \textbf{Real-R}, and \textbf{Real-S}, respectively.

\begin{itemize}[leftmargin=*]
    \item\textbf{TPC-H}: Compared to DTA, GPT-5 performs similarly or better for most queries. For example, it significantly outperforms DTA for queries 5 and 17, whereas DTA is advantageous for queries 14 and 21.
    Since \textbf{TPC-H} is a well-known benchmark that is publicly available, it may have been seen during GPT-5's training.
    Therefore, we also tried to ``anonymize'' the queries by replacing the table and column names with randomly generated identifiers. 
    However, we observed similar performance for GPT-5 on the anonymized version, which is consistent with the findings reported in prior work~\cite{GiannakourisT25}.
    \item\textbf{Real-D}: We observe that LLM significantly outperforms DTA for several queries (e.g., queries 1, 4, 9, 15, 18, 20, 27, and 31), while DTA performs better for others (e.g., queries 3, 11, 22, 23, 26, 28, and 30). 
    Overall, the observations are similar to those from \textbf{TPC-H}, despite the substantially higher schema and query complexity of \textbf{Real-D}.
    \item\textbf{Real-M}: 
    DTA achieves better performance for only a small number of queries (e.g., queries 7 and 29), whereas LLM outperforms DTA for many others (e.g., queries 6, 9, 12, 19, 27, and 33). Moreover, for several queries (e.g., queries 6, 27, and 33), indexes recommended by DTA lead to query performance regressions (QPRs)~\cite{DingDWCN18,DingDM0CN19,Wu25,WuDXNC25}. In these cases, LLM is able to identify a better plan that improves query execution time.
    \item\textbf{Real-R}: The observations are similar to those from \textbf{Real-M}. Except for a few queries (e.g., queries 6, 10, and 25), LLM outperforms DTA, sometimes quite significantly (e.g., for queries 5 and 22).
    Notably, DTA's recommendation results in a severe QPR for query 22 (nearly a $10\times$ slowdown in execution time), whereas LLM improves by approximately 60\%, reducing its execution time from 10 seconds to 4 seconds.
    \item\textbf{Real-S}: LLM and DTA perform similarly for \textbf{Real-S} queries. DTA significantly outperforms LLM for queries 2 and 8, while LLM significantly outperforms DTA for query 7. We do not observe QPRs for the \textbf{Real-S} queries.
\end{itemize}

Based on the observations, we have the following conclusion.

\begin{observation}
With a small number of trials, the best outcome of LLM-driven index tuning underperforms, matches, or substantially outperforms DTA across different sets of queries, each comprising a considerable fraction of our test cases. 
\end{observation}

This observation highlights the complementary potential of LLM-driven index tuning for industrial single-query workloads, as it can identify configurations that substantially outperform those produced by DTA for many queries, particularly in cases where DTA's recommendation leads to a performance regression.

\begin{figure*}
\centering\vspace{-1em}
\setcounter{subfigure}{0}
\subfigure[\textbf{TPC-H}]{ \label{fig:index-usage:tpch}
    \includegraphics[width=0.47\textwidth]{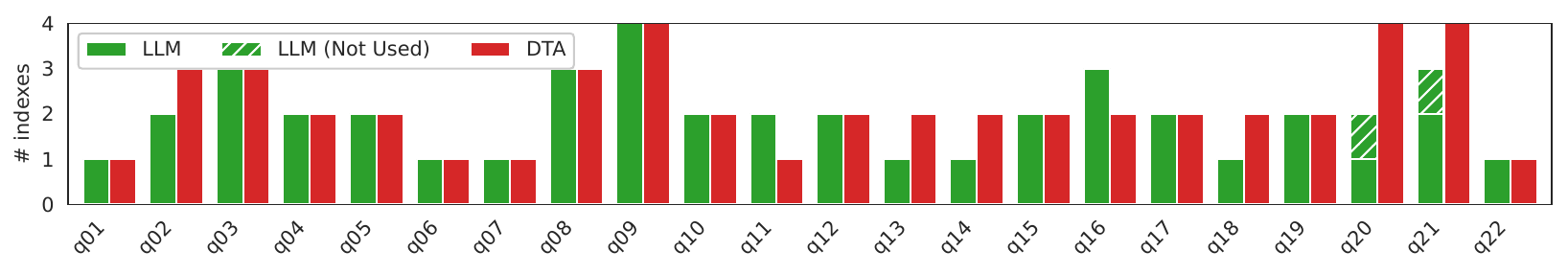}}
\setcounter{subfigure}{3}    
\subfigure[\textbf{Real-R}]{ \label{fig:index-usage:real-r}
    \includegraphics[width=0.51\textwidth]{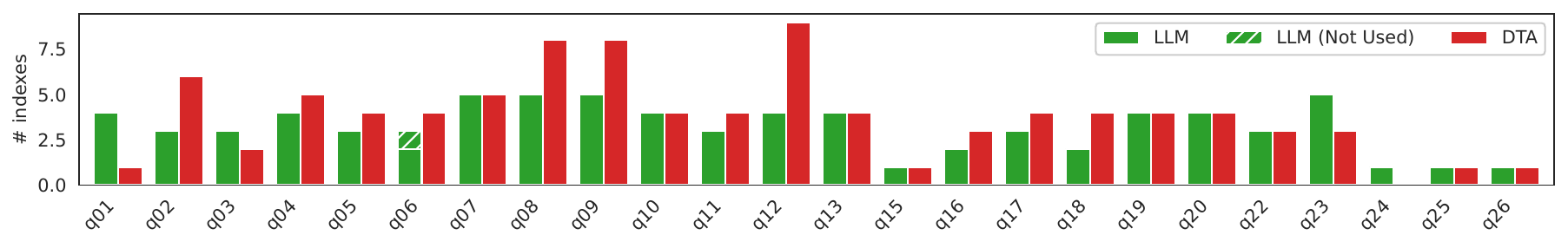}}
\setcounter{subfigure}{1}
\subfigure[\textbf{Real-D}]{ \label{fig:index-usage:real-d}
    \includegraphics[width=0.69\textwidth]{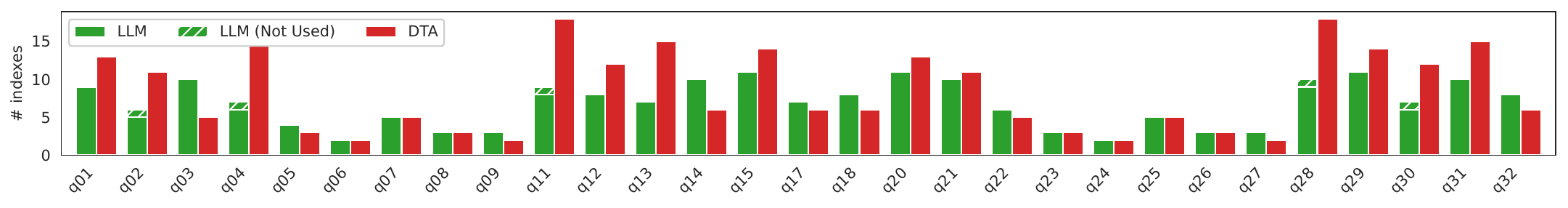}}
\setcounter{subfigure}{4}
\subfigure[\textbf{Real-S}]{ \label{fig:index-usage:real-s}
    \includegraphics[width=0.28\textwidth]{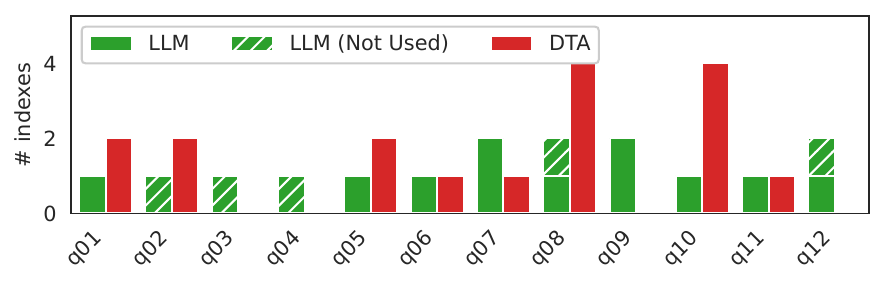}}
\setcounter{subfigure}{2}
\subfigure[\textbf{Real-M}]{ \label{fig:index-usage:real-m}
    \includegraphics[width=0.98\textwidth]{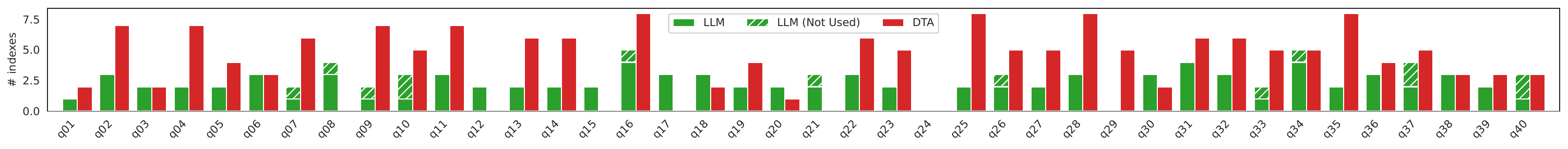}}
\vspace{-1.5em}
\caption{Comparison of index usage between LLM-driven index tuning and DTA for tuning single-query workloads.}
\label{fig:index-usage}
\vspace{-1.5em}
\end{figure*}

\vspace{-0.5em}
\subsection{Efficacy of LLM-Recommended Indexes}
\label{sec:single-query-index-usage}

Since we impose no explicit constraints on the number of indexes recommended by LLM, its strong performance may stem from proposing a sufficiently large set of indexes, thus increasing the likelihood of covering optimal configurations. 
To assess this possibility, we conduct a drill-down analysis to better understand the real efficacy of LLM-recommended indexes, as reported in Section~\ref{sec:single-query-overview}. 

Figure~\ref{fig:index-usage} presents the number of LLM-recommended indexes utilized by query plans for single‑query workloads. 
For comparison, we also show the number of indexes recommended by DTA.



\begin{itemize}[leftmargin=*]
    \item\textbf{TPC-H}: For most queries, all indexes recommended by LLM are utilized by their query plans, with only a couple exceptions (e.g., queries 20 and 21). Interestingly, DTA recommends more indexes for certain queries (e.g., queries 2, 13, 14, 18, 20, and 21), which are all utilized by the corresponding query plans.
    \item\textbf{Real-D}: Most of the indexes recommended by LLM are utilized by the corresponding query execution plans. For more than half of the queries, DTA recommends substantially more indexes compared to LLM. For example, DTA recommends 17 indexes for the query 4 whereas LLM recommends only 7 indexes.
    \item\textbf{Real-M}: Similarly, DTA typically recommends considerably more indexes than LLM. However, it does not recommend any indexes for several queries (e.g., queries 8, 12, 15, 17, 21, and 24). Meanwhile, LLM recommends fewer indexes, most of which are utilized by the corresponding query execution plans.
    \item\textbf{Real-R}: DTA and LLM recommend similar numbers of indexes for most queries, with a few exceptions (e.g., queries 8, 9, and 12) where DTA recommends significantly more indexes.
    \item\textbf{Real-S}: Indexes recommended by LLM are not utilized by the corresponding query plans for several queries (e.g., queries 2, 3, and 4). Meanwhile, DTA does not recommend any indexes for quite a few queries (e.g., queries 3, 4, 9, and 12). 
\end{itemize}



Based on the above observations, we conclude the following.
\begin{observation}
Most of the indexes that LLM proposes for single queries are effectively utilized by the corresponding query plans. In many cases, LLM recommends even fewer indexes than DTA.
\end{observation}



\vspace{-0.5em}
\subsection{What Happens When LLM Performs Better?}
\label{sec:single-query-estcost}





\begin{figure*}
\centering\vspace{-1em}
\subfigure[\textbf{TPC-H}]{ \label{fig:index-benefit:tpch}
    \includegraphics[width=0.19\textwidth]{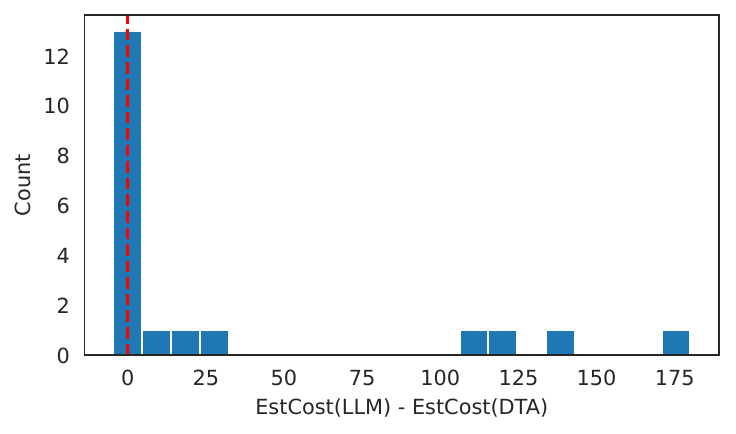}}
\subfigure[\textbf{Real-D}]{ \label{fig:index-benefit:real-d}
    \includegraphics[width=0.19\textwidth]{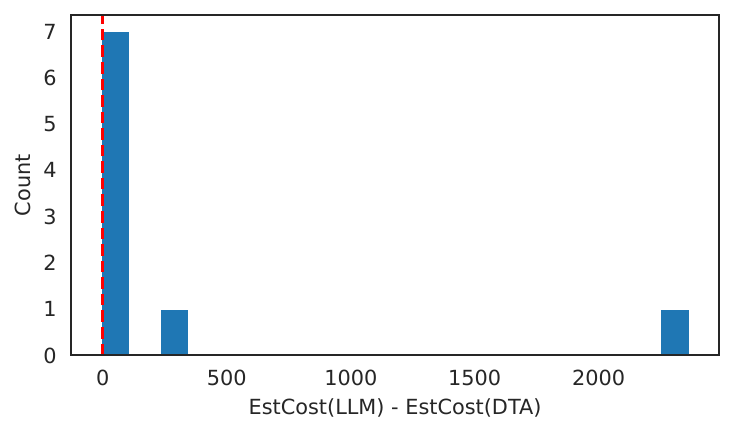}}
\subfigure[\textbf{Real-M}]{ \label{fig:index-benefit:real-m}
    \includegraphics[width=0.19\textwidth]{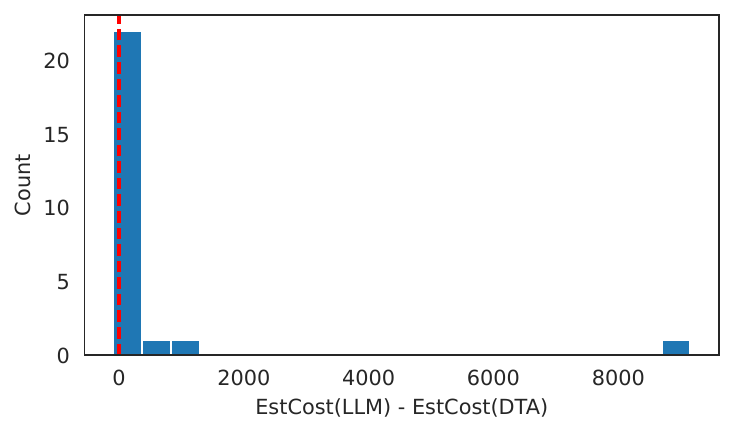}}
\subfigure[\textbf{Real-R}]{ \label{fig:index-benefit:real-r}
    \includegraphics[width=0.19\textwidth]{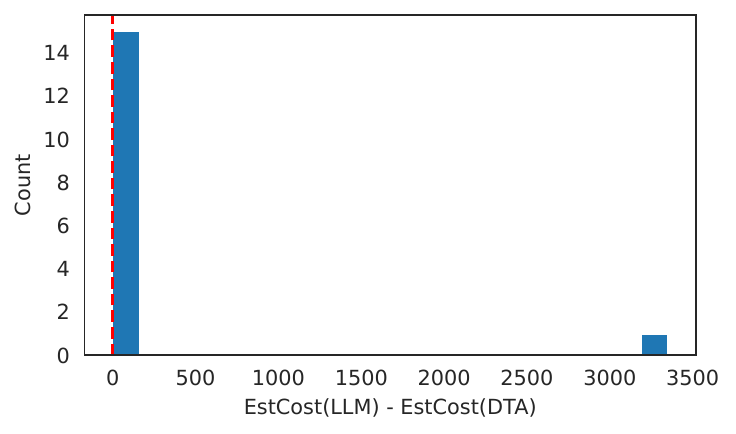}}
\subfigure[\textbf{Real-S}]{ \label{fig:index-benefit:real-s}
    \includegraphics[width=0.19\textwidth]{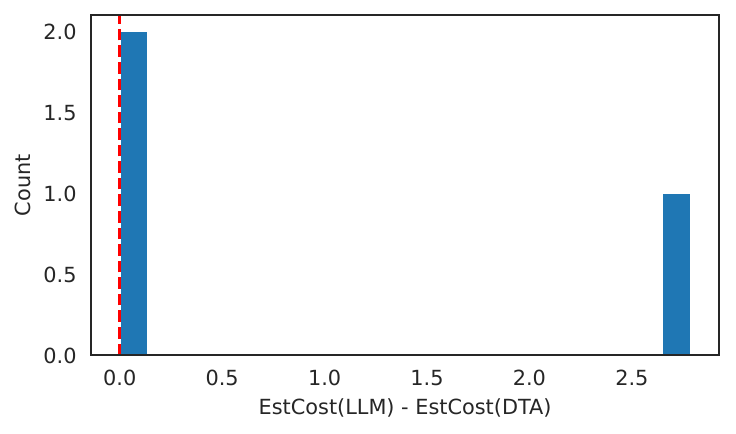}}
\vspace{-1.5em}
\caption{Comparison of the estimated costs between LLM‑driven index tuning and DTA for single‑query workloads.} 
\label{fig:index-benefit}
\vspace{-1.5em}
\end{figure*}

Next, we conduct a deeper investigation into the cases where LLM outperforms DTA. Specifically, we examine these queries by comparing the estimated costs of query plans produced using indexes recommended by DTA and LLM. Figure~\ref{fig:index-benefit} summarizes the results. 
Only cases where LLM outperforms DTA are shown.

Each graph in Figure~\ref{fig:index-benefit} is a histogram showing the distribution of queries by the difference in estimated costs between the plans produced by LLM and DTA. 
The $x$-axis denotes this cost difference: negative values indicate that the LLM‑recommended indexes are favored by query optimizer’s cost model, while positive values indicate the opposite. 
The red dashed line indicates zero difference. 
The $y$‑axis represents the number of queries that fall into each cost‑difference bin. 
We have the following observation.

\begin{observation}
For queries where LLM‑recommended indexes yield faster execution time, their optimizer‑estimated costs are consistently higher than those of the indexes recommended by DTA.
\end{observation}


This reveals the deployment risk shared by DTA and other cost‑based index tuners.
Although these cost-based index tuners 
are widely adopted in industrial applications~\cite{DasGIJJNRSXC19,ValentinZZLS00,dexter-2,ChakkappenKMKSZZLZ25}, their efficacy largely depends on the accuracy of the model used by the underlying query optimizer. 
It is well known that query optimizer's cost estimates can be inaccurate for various reasons, such as errors from cardinality estimation~\cite{WangQWWZ21,LeisGMBK015,WuNS16,IoannidisC91}.
Therefore, while these cost-based index tuners are generally reliable, they can sometimes result in poor index recommendations. 

In contrast, the index tuning approach taken by LLM is based on reasonable rule-of-thumb heuristics (Section~\ref{sec:llm-efficacy}) derived from the world knowledge acquired during LLM's training. 
Such heuristics may not be comprehensive enough to cover all plan transformation rules built into a modern cost-based query optimizer~\cite{Cascades}, but they can be used to improve query execution.
More importantly, although LLM can miss certain opportunities that DTA can find (e.g., when LLM underperforms DTA in Figure~\ref{fig:llm-vs-dta}), its heuristic-based recommendation is \emph{less susceptible to inaccurate cost estimates}. 
Table~\ref{tab:cost-vs-time} presents such examples, where query plans using LLM-recommended indexes have a higher estimated cost but achieve a substantially better execution time.

Due to inaccurate cost estimates, DTA would not select the configurations identified by LLM since their estimated costs are larger (than the ones finally selected by DTA), even when they fall within the search space of DTA. Such inaccuracies can sometimes lead to severe QPRs, such as query 4 from \textbf{Real-D} and query 27 from \textbf{Real-M}. 
These results help explain the complementary performance of LLM-driven index tuning relative to DTA.

\vspace{-0.5em}
\subsection{Performance Variance of LLM}
\label{sec:single-query-workloads:robustness}



So far, we have focused on the best results from LLM, highlighting its strong performance potential. 
However, we also observe substantial performance variance in LLM-recommended configurations. 
To further illustrate this, this section examines the \emph{worst} outcomes among the five invocations, indicated by the star markers in Figure~\ref{fig:llm-vs-dta}, and compares them against the best results.


\begin{itemize}[leftmargin=*]
    \item\textbf{TPC-H}: The gap between the best and worst outcomes across the five LLM invocations is substantial for many queries (e.g., queries 1, 3, 4, 5, 8, 9, 12, and 20). 
    This gap is particularly pronounced for query 20, where the worst outcome results in a severe QPR. With the worst-case outcomes, LLM no longer outperforms DTA for any query and, in most cases, performs substantially worse.
    \item\textbf{Real-D}: There are a few cases where the worst LLM response achieves a similar level of improvement to the best one and still outperforms DTA (e.g., queries 1, 9, 20, and 27). However, for the majority queries, the worst response offers no improvement in execution time. For some queries (e.g., queries 13, 14, and 31), it leads to QPRs and eventually timeouts, whereas the corresponding best response does not exhibit such behavior.
    \item\textbf{Real-M}: The execution time variance across different LLM invocations remains substantial. The worst responses lead to QPRs for queries such as 5 and 27. Moreover, they consistently underperform DTA, even for queries where the best LLM outcomes outperform DTA (e.g., queries 3, 9, 19, 27 and 34).
    \item\textbf{Real-R}: We observe two significant QPRs caused by the worst LLM responses, both resulting in timeouts. Although these worst-case outcomes can still accelerate some of the queries, they outperform DTA for only two queries (i.e., queries 22 and 24).
    \item\textbf{Real-S}: The worst of the five LLM responses cannot match the performance of DTA. In particular, for query 7, which is the only case where the best LLM response outperforms DTA, the worst response is unable to produce any improvement.
\end{itemize}

\begin{observation}
LLM-driven index tuning exhibits substantial performance variance within a small number of trials. Its worst outcomes may significantly underperform DTA and, in many cases, lead to severe performance regressions.
\end{observation}


This observation motivates careful performance validation before adopting LLM-recommended indexes in production. We discuss this aspect in more detail in Section~\ref{sec:lesson:cost}.




\begin{table}[t]
\vspace{0.5em}
\caption{Examples of inaccurate estimated cost with the what-if API vs. actual plan execution time.} 
\vspace{-1em}
\label{tab:cost-vs-time}
\small
\centering
\begin{tabular}{|l|r|r|r|r|r|}
\hline
Query & DTA Cost & DTA Time & GPT-5 Cost & GPT-5 Time \\
\hline
\hline
TPC-H q05	&	132.736	&	3.437s	&	268.35	&	0.603s	\\
\hline
Real-D q04	&	1402.38	&	134.324s	&	3767.51	&	19.211s	\\
Real-D q27	&	268.087	&	13.427s	&	295.872	&	4.375s	\\
\hline
Real-M q19	&	2934.32	&	12.522s	&	12088.7	&	1.549s	\\
Real-M q27	&	553.253	&	10.210s	&	1643.99	&	1.202s	\\
\hline
Real-R q22	&	34.3663	&	111.259s	&	40.7258	&	4.268s	\\
\hline
Real-S q07	&	5.64505	&	11.398s	&	8.43237	&	3.168s	\\
\hline
\end{tabular}
\end{table}
\vspace{-0.5em}
\section{Distilling Insights from LLM}
\label{sec:llm-efficacy}


\cut{
The effectiveness of LLM-driven index tuning for single-query workloads is inspiring.
However, its instability (Section~\ref{sec:single-query-workloads:robustness}) raises the question of whether this efficacy reflects a genuine understanding of the index tuning problem or arises primarily by chance.

In this section, we investigate this question in detail.
We start by examining the underlying reasoning processes of GPT-5 for tuning single-query workloads. 
We observe that GPT-5's reasoning follows several intuitive insights that align with human judgment and can be summarized into a few rules of thumb. Based on this observation, we explore whether these insights can be distilled into a simple rule-based index tuner and evaluate its efficacy.
}

We found the effectiveness of LLM-recommended indexes on real-world single-query workloads surprising (Section~\ref{sec:single-query-workloads}). However, we also found that the non-determinism of LLM leads to significant variance (Section~\ref{sec:single-query-workloads:robustness}) in the index recommendation quality for the same query as well as across queries. We therefore conducted two follow-up studies. First, we analyzed the \emph{reasoning process} output by the LLM during inferencing to characterize the basis on which LLM generates its recommendations. Our analysis reveals several rules of thumb used by the LLM when recommending indexes (Section~\ref{sec:llm-efficacy:reasoning}). Second, using these rules of thumb, we built a simple \emph{deterministic} index tuner. Our intent was to evaluate whether such an approach could derive most of the benefits of an LLM without inherting its variance in quality (Section~\ref{sec:llm-efficacy:simple}).

\vspace{-1em}
\subsection{Analysis of LLM's Reasoning Process}\label{sec:llm-efficacy:reasoning}

\begin{figure}[t]
\centering
    \vspace{1em}
    \includegraphics[clip, trim=1.5cm 2cm 1.3cm 4cm, width=\columnwidth]{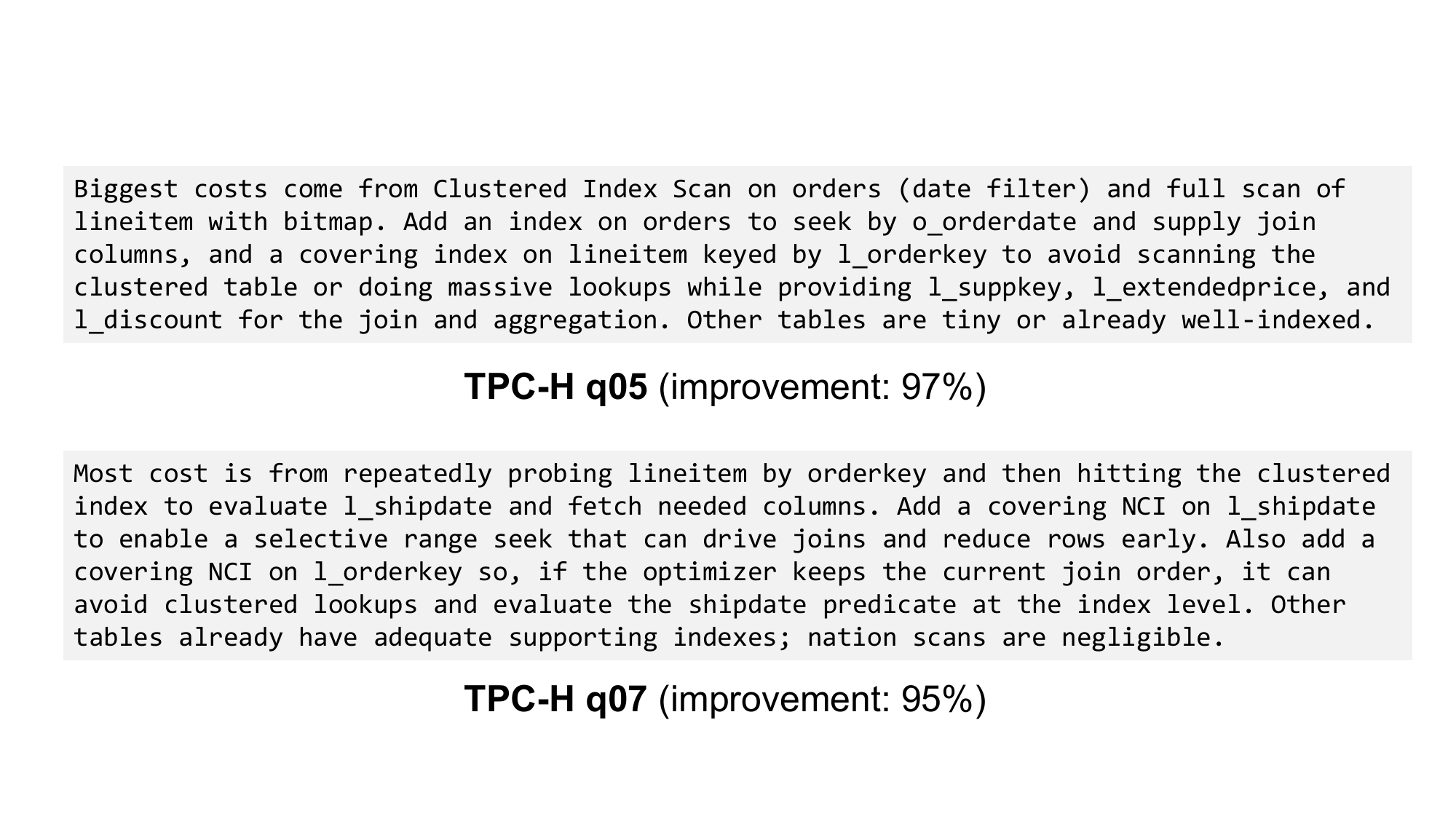}
\vspace{-2.5em}
\caption{Examples of GPT-5's reasoning process.} 
\label{fig:reasoning-example}
\end{figure}

To analyze GPT-5's reasoning process, we instruct it to output its reasoning together with the recommended indexes. Figure~\ref{fig:reasoning-example} presents two typical examples. 
The recommended configurations result in a 97\% improvement for \textbf{TPC-H} query 5 and a 95\% improvement for query 7. 
We observe that LLM's reasoning usually begins with identifying the \emph{bottleneck} operators in the input query plan, followed by analysis of how to use indexes to improve these bottlenecks. 
For example, for query 5, GPT-5 argues that having an index built on the \textsf{orders} table to seek by \textsf{o\_orderdate} can replace the costly \textsf{Clustered Index Scan} operator and lead to significant speedup. Examining these LLM reasoning processes during our evaluation, we make the following observation. 

\begin{observation}
In summary, GPT‑5’s reasoning reveals several simple rules of thumb: (1) prioritize building indexes that reduce costly scan operations; (2) order index key columns according to cues from the query plan, such as filters, joins, and aggregates; (3) introduce covering indexes when possible to enable index‑only scans and avoid expensive full table scans; and (4) ignore small table scans.
\end{observation}




These rules of thumb are much simpler than DTA’s approach to index recommendation. 
DTA takes a cost-based approach that relies on the what-if API provided by the query optimizer. 
Even during candidate index generation, it makes cost-based decisions~\cite{dta}. 
For example, after parsing the workload queries, DTA uses a \emph{frequent pattern mining} procedure~\cite{AgrawalCN01} to identify \emph{interesting} table subsets and column groups based on their frequency and costs. 
Multi-column indexes are restricted to those defined on interesting column groups. 
For indexes on columns with filter conditions, DTA orders the columns by the estimated selectivity of the corresponding predicates. 
Moreover, DTA uses \emph{index merging}~\cite{ChaudhuriN99} to combine similar candidate indexes, which is again guided by cost estimates.

\begin{figure*}
\centering\vspace{-1em}
\subfigure[\textbf{TPC-H} $\alpha=0$]{ \label{fig:simple-tuner:tpch}
    \includegraphics[width=0.19\textwidth]{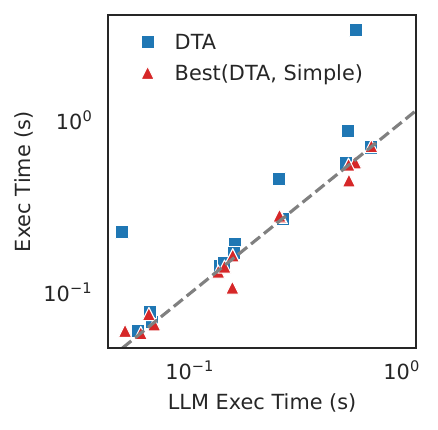}}
\subfigure[\textbf{Real-D} $\alpha=0$]{ \label{fig:simple-tuner:real-d}
    \includegraphics[width=0.19\textwidth]{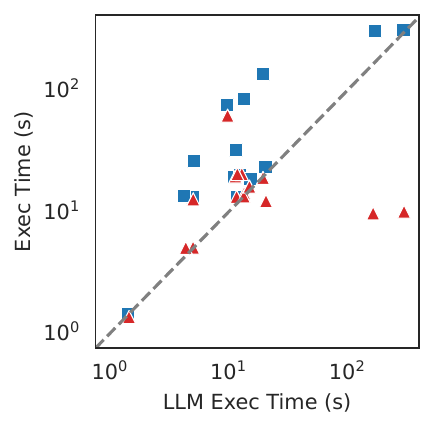}}
\subfigure[\textbf{Real-M} $\alpha=0$]{ \label{fig:simple-tuner:real-m}
    \includegraphics[width=0.19\textwidth]{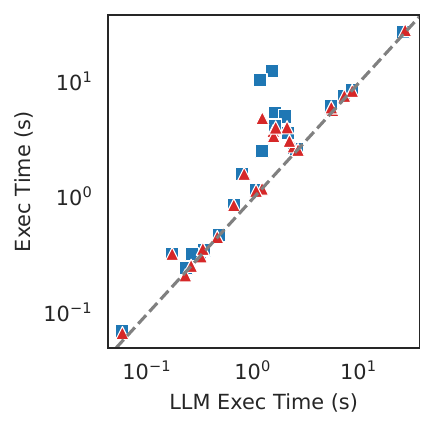}}
\subfigure[\textbf{Real-R} $\alpha=0$]{ \label{fig:simple-tuner:real-r}
    \includegraphics[width=0.19\textwidth]{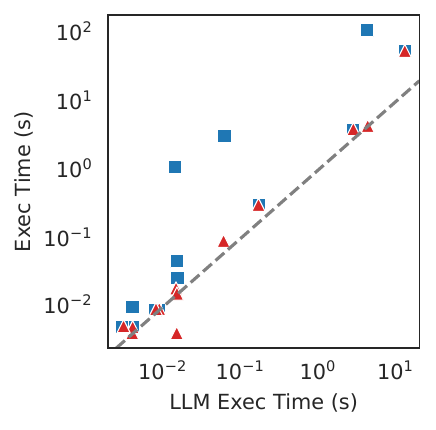}}
\subfigure[\textbf{Real-S} $\alpha=0$]{ \label{fig:simple-tuner:real-s}
    \includegraphics[width=0.19\textwidth]{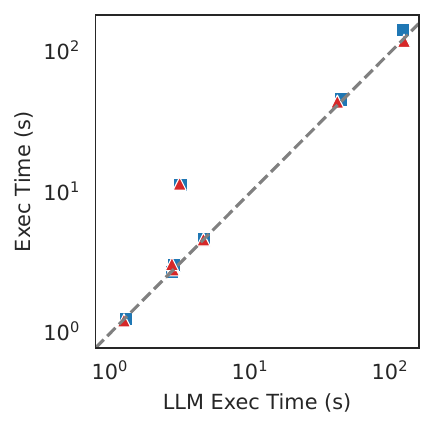}}
\subfigure[\textbf{TPC-H} $\alpha=0.01$]{ \label{fig:simple-tuner:tpch:0.01}
    \includegraphics[width=0.19\textwidth]{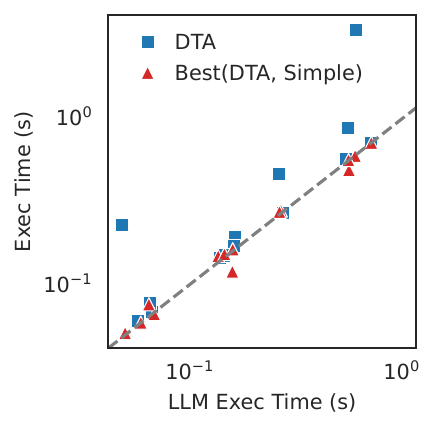}}
\subfigure[\textbf{Real-D} $\alpha=0.01$]{ \label{fig:simple-tuner:real-d:0.01}
    \includegraphics[width=0.19\textwidth]{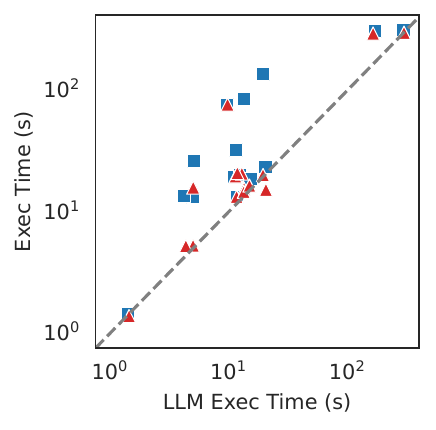}}
\subfigure[\textbf{Real-M} $\alpha=0.01$]{ \label{fig:simple-tuner:real-m:0.01}
    \includegraphics[width=0.19\textwidth]{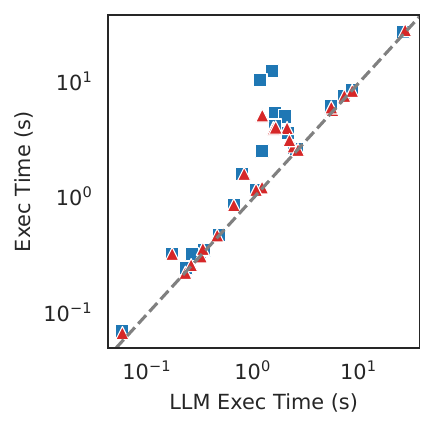}}
\subfigure[\textbf{Real-R} $\alpha=0.01$]{ \label{fig:simple-tuner:real-r:0.01}
    \includegraphics[width=0.19\textwidth]{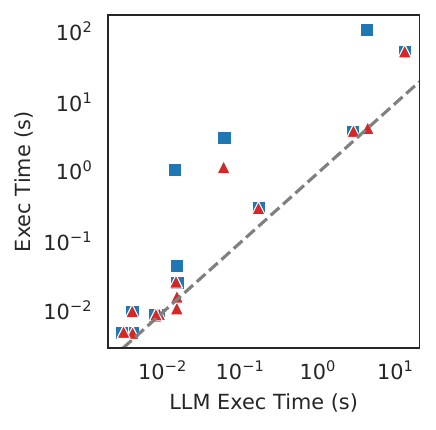}}
\subfigure[\textbf{Real-S} $\alpha=0.01$]{ \label{fig:simple-tuner:real-s:0.01}
    \includegraphics[width=0.19\textwidth]{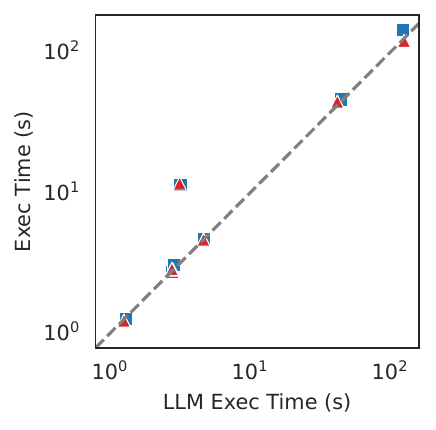}}
\vspace{-1em}
\caption{Evaluation of the simple index tuner (y-axis: $\blacksquare$ DTA time, $\blacktriangle$ best time of DTA and the simple index tuner).}
\label{fig:simple-tuner}
\vspace{-1.5em}
\end{figure*}

\vspace{0.5em}
\subsection{A Simple Rule-based Index Tuner}\label{sec:llm-efficacy:simple}

These rules of thumb that we observed from GPT-5's reasoning process align with human intuition. Together with the strong but unstable performance that LLM shows in Section~\ref{sec:single-query-workloads}, it will be interesting to ask whether we can \emph{distill some insights from LLM into a deterministic program}. 
To explore this direction, we develop a simple rule-based index tuner guided by these summarized insights.


The simple tuner constructs one covering index per table by analyzing only the original query plan. 
A technical problem is to determine the order of \emph{key columns} for each index, as we do not observe a consistent pattern from GPT-5. 
Our approach follows the implicit column-access order with respect to selective operators in the query plan. 
For example, if the plan first filters a table on column $A$, then joins it with another table on column $B$, and the final output is ordered by column $C$, we construct an index with key columns ordered as $(A, B, C)$. Any additional referenced columns from the same table (e.g., those needed by projection) are incorporated into the index as \emph{included columns}. 
This approach is motivated by two considerations. 
First, 
the index thus generated is likely to be used to improve the current plan, which is the best plan returned by the query optimizer with existing indexes. 
Second, the new index is less likely to significantly change the current join order, reducing the risk of selecting a new but potentially disastrous one~\cite{WuDXNC25}.

Algorithm~\ref{alg:simple} presents the details of our rule-based index tuner.
It begins by enumerating all base tables referenced in the plan and initializing three variables for each table: (1) the ordered key‑column list $K$, (2) the set of referenced columns $A$, and (3) the cumulative access cost $C$ (lines 2 to 3). It then performs a depth‑first traversal of the query plan via the procedure \textsf{Traverse}, which updates $A$ and $C$ whenever a \textsf{Scan} operator is encountered (lines 14 to 16), and appends columns to $K$ when they appear in selective or ordering operators, e.g., \textsf{IndexSeek}, \textsf{Filter}, \textsf{Join}, \textsf{GroupBy}, \textsf{OrderBy} (lines 17 to 18). After traversing the query plan, it constructs a covering index for each table whose cumulative access cost is significant enough, using the collected list of key columns and the set of referenced columns (lines 6 to 8).
To determine the significance of the cumulative access cost of a table, we specify a percentage threshold $0\leq\alpha<1$.
A table is costly enough if its cumulative access cost is greater than $\alpha$ timed by the total plan cost (line 7).

\begin{algorithm}[t]
\footnotesize
\SetAlgoLined
\KwIn{$r$: root of the plan tree; $\alpha$: index-building threshold.}
\KwOut{$\mathcal{I}$: set of index recommendations.}
$\mathcal{T} \leftarrow \texttt{getAllBaseTables}(r)$\;
\ForEach{$t \in \mathcal{T}$}{
  $K[t] \leftarrow [\,]$; $A[t] \leftarrow \emptyset$; $C[t] \leftarrow 0$\;
}
\texttt{Traverse}($r$, $K$, $A$, $C$)\;

$\mathcal{I} \leftarrow \emptyset$\;
\ForEach{$t \in \mathcal{T}$}{
  \uIf{$C[t] > \alpha \cdot r.\mathrm{cost}$}{
    \tcp{Build a covering index.}
    $\mathcal{I} \leftarrow \mathcal{I} \cup \texttt{buildIndex}(\textsf{table}=t,~\textsf{keyCols}=K[t],~\textsf{includedCols}=A[t]\setminus K[t])$\;
  }
}

\Return{$\mathcal{I}$;}

\quad\\
\SetKwProg{Fn}{\underline{\textbf{Traverse}}}{}{}
\Fn{($n$, $K$, $A$, $C$)}{
    \ForEach{$n_c \in n.\mathrm{children}$}{
        \texttt{Traverse}($n_c$, $K$, $A$, $C$)\;
    }
    \uIf{$n.\mathrm{type} = \textsf{Scan}$}{
        \tcp{Accumulate base table access cost.}
        $C[n.\mathrm{table}] \leftarrow C[n.\mathrm{table}] + n.\mathrm{cost}$\; 
        \tcp{Collect all referenced columns.}
        $A[n.\mathrm{table}] \leftarrow A[n.\mathrm{table}] \cup n.\mathrm{allRefCols}$\; 
    }
    \ForEach{$c \in n.\mathrm{seekCols} \cup n.\mathrm{predicateCols} \cup n.\mathrm{groupByCols} \cup n.\mathrm{orderByCols} \cup n.\mathrm{joinKeyCols}$}{
        \tcp{Collect key columns in order of first access.}
        $K[c.\mathrm{table}] \leftarrow \texttt{appendIfNotExists}(K[c.\mathrm{table}], c)$\;
    }
}
\caption{\texttt{SimpleIndexRecommendation}($r$, $\alpha$)}
\label{alg:simple}
\end{algorithm}


\vspace{-0.5em}
\subsection{Evaluation of the Rule-based Index Tuner}\label{sec:llm-efficacy:simple_eval}
Figure~\ref{fig:simple-tuner} presents the evaluation results of the rule-based index tuner (Algorithm~\ref{alg:simple}), with all five workloads using $\alpha=0$ (i.e., building indexes for all tables) and $\alpha=0.01$ (i.e., skipping tables whose estimated access cost $\leq1\%$ of the total plan cost). 
Again, only cases where LLM outperforms DTA are shown.
To highlight the complementary benefit that this simple tuner can offer, we report results only for queries where the LLM outperforms DTA. 

For each query, we compare the execution time obtained under different tuning strategies. The x-axis shows the best execution time achieved by LLM among its five responses. The y-axis reports the execution time under two interpretations, distinguished by the shape of the marker: square markers denote the execution time under DTA, while triangle markers denote the better execution time between DTA and the simple tuner. 
The latter reflects the best possible performance when combining the recommendations of both DTA and the simple rule-based tuner. 
The dashed line indicates equal performance on both axes.

\begin{figure*}[t]
\centering\vspace{-0.5em}
\subfigure[\textbf{TPC-H}]{ \label{fig:multi-query:tpch}
    \includegraphics[width=0.32\textwidth]{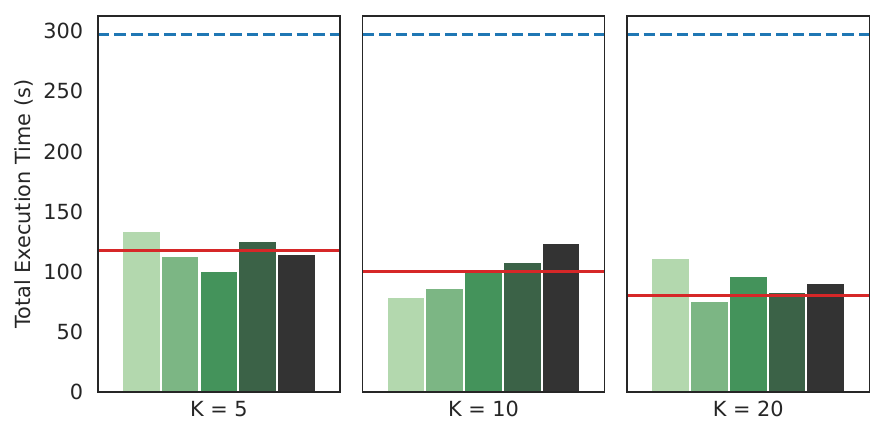}}
\subfigure[\textbf{Real-D}]{ \label{fig:multi-query:real-d}
    \includegraphics[width=0.32\textwidth]{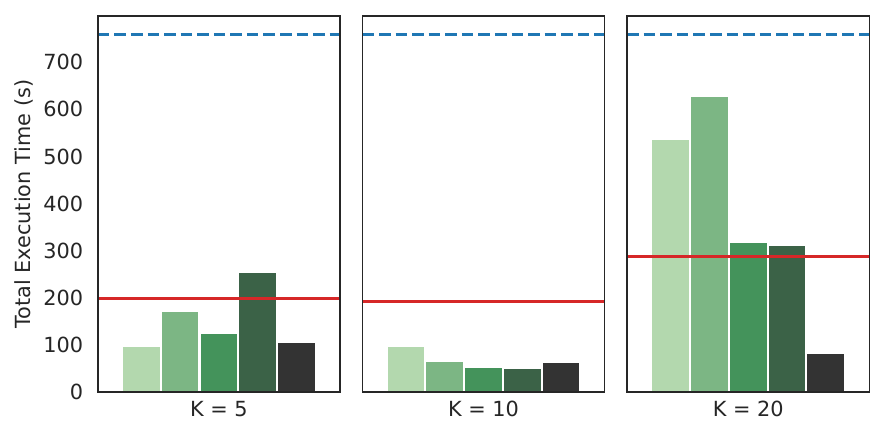}}
\subfigure[\textbf{Real-M}]{ \label{fig:multi-query:real-m}
    \includegraphics[width=0.32\textwidth]{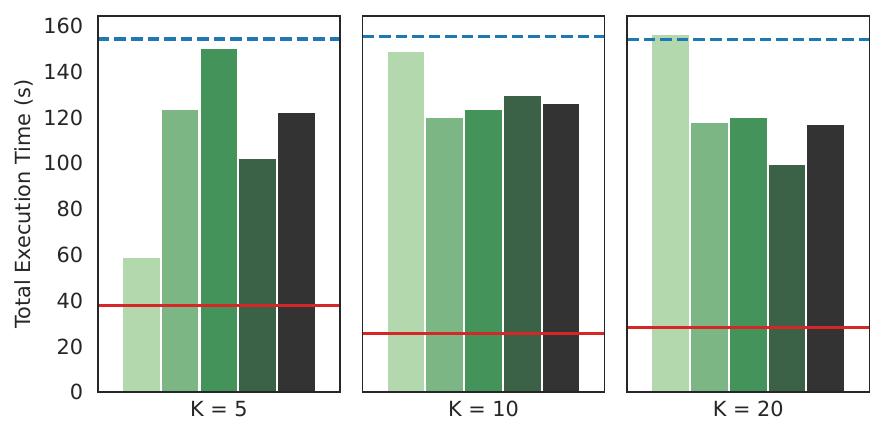}}
\subfigure[\textbf{Real-R}]{ \label{fig:multi-query:real-r}
    \includegraphics[width=0.32\textwidth]{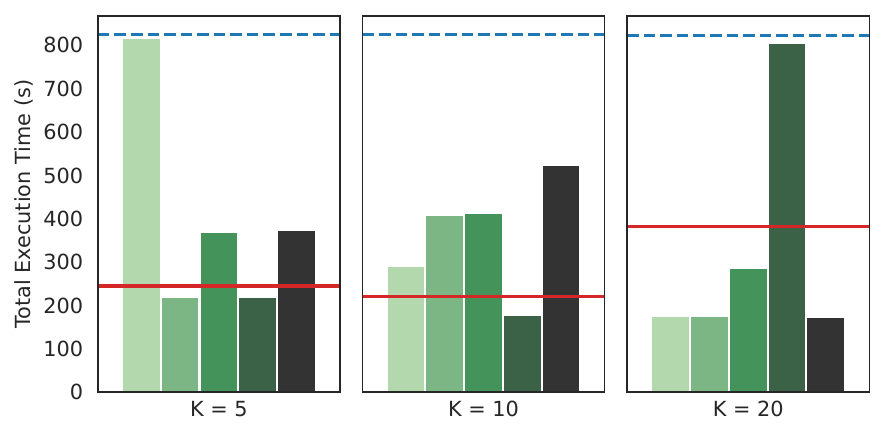}}
\subfigure[\textbf{Real-S}]{ \label{fig:multi-query:real-s}
    \includegraphics[width=0.32\textwidth]{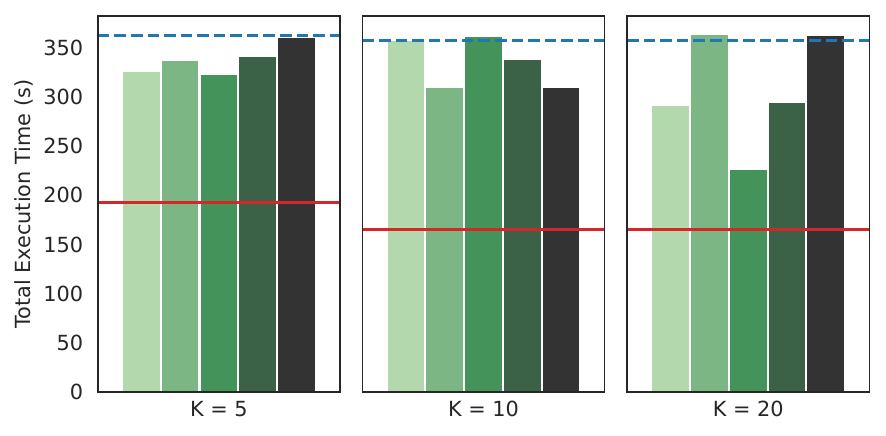}}
\vspace{-1.5em}
\caption{LLM-driven index tuning (five invocations shown as bars) vs. DTA (red line) for tuning multi-query workloads.}
\label{fig:multi-query}
\vspace{-1.5em}
\end{figure*}

\begin{itemize}[leftmargin=*]
    \item\textbf{TPC-H}: There are four queries where the best among the five LLM responses substantially outperforms DTA, as indicated by the four blue squares far above the diagonal line. In all of these cases, the simple rule-based tuner (with both $\alpha=0$ and $0.01$) identifies recommendations that are comparable to those of the LLM, with all red triangles lying close to the diagonal.
    \item\textbf{Real-D}: With $\alpha=0$, the simple rule-based tuner effectively reduces the number of queries for which DTA performs substantially worse than LLM. In particular, two queries that originally timed out (i.e., exceeding the five-minute limit) or nearly timed out after tuning by both LLM and DTA can now be completed in about ten seconds, as indicated by the two rightmost pairs of markers. In contrast, skipping indexes on low-cost tables (with $\alpha = 0.01$) does not improve these cases, indicating that even small tables can matter from an indexing perspective. Aside from these two queries, the simple tuner yields similar improvements across different threshold values. However, there are still queries where the LLM finds significantly better configurations even after incorporating the rule-based tuner, as reflected by triangles above the diagonal line.
    \item\textbf{Real-M}: The rule-based tuner substantially narrows the gap between DTA and LLM for queries with the largest performance differences (note that the execution time is shown with a logarithmic scale). However, it does not produce comparable recommendations in all cases. 
    The performance of the rule-based tuner is also insensitive to the choice of threshold $\alpha$.
    \item\textbf{Real-R}: The observation is similar to that with \textbf{Real-M}, as the rule-based tuner (with both $\alpha=0$ and $0.01$) can improve DTA and narrow its gap with LLM. Nevertheless, not all queries can be improved to match the performance of LLM.
    \item\textbf{Real-S}: There is only one query for which LLM substantially outperforms DTA. 
    Unfortunately, the rule-based tuner cannot identify recommendations that improve DTA for this case.
\end{itemize}

\begin{observation}
In cases where DTA underperforms LLM, the simple rule-based index tuner produces better recommendations than DTA in approximately 60\% of our test queries, and achieves performance comparable to or better than LLM in about 40\% of all test cases.
\end{observation}

This observation is encouraging, as it suggests that, by distilling insights from LLM, we can achieve substantial improvements in many cases by only validating two sets of recommendations: those from DTA and those from the simple rule-based index tuner. 


\vspace{-0.5em}
\paragraph*{Remarks} 
The simple tuner serves as a proof of concept demonstrating that effective heuristics can be distilled from LLM without inheriting the risks of high variance in index recommendation quality. 
\revision{We observe that this simple rule-based tuner is effective on its own, particularly in cases where DTA's recommendations underperform. A more comprehensive evaluation of the rule-based tuner is included in the full version of this paper~\cite{full-version}.}

\section{Multi-query Workloads}
\label{sec:multi-query-workloads}

We extend our study to multi-query workloads, a more common setting in industrial practice.
Unlike tuning a single query, tuning a multi-query workload calls for the index tuner to disregard indexes that are promising locally (for some queries) but unfavorable globally (for the entire workload).
Practical constraints, such as the number of indexes or storage space limits, further restrict the ability to materialize all beneficial indexes. 

\subsection{Overview of Evaluation Results}
\label{sec:multi-query-workloads:overview}

We construct multi-query workloads as follows. For \textbf{TPC-H}, \textbf{Real-R}, and \textbf{Real-S}, we include all queries in a single workload. For \textbf{Real-D} and \textbf{Real-M}, whose queries are more complex and exceed GPT-5’s context window when considered jointly, we select the 10 queries with the largest observed improvements under single-query tuning by GPT-5 (as presented in Section~\ref{sec:single-query-workloads}).

Figure~\ref{fig:multi-query} presents the results for five invocations on all multi-query workloads, with the blue dashed line denoting the workload execution time with the original index configuration. 
The solid red line represents the workload execution time with the indexes recommended by DTA.
The average response time of GPT-5 is 4.4, 3.8, 2.3, 2.4, and 3.4 minutes for \textbf{TPC-H}, \textbf{Real-M}, \textbf{Real-D}, \textbf{Real-R}, and \textbf{Real-S}, respectively. 
We vary the constraint $K$ on the number of indexes allowed as $K\in\{5, 10, 20\}$.
We do not observe violation of these constraints from the responses of LLM in our evaluation.


\begin{itemize}[leftmargin=*]
    \item\textbf{TPC-H}: The performance of LLM is similar to that of DTA, regardless of the maximum number of indexes allowed $K$.
    \item\textbf{Real-D}: 
    With $K=5$ or $K=10$, LLM-recommended indexes yield lower execution time than DTA-recommended indexes.
    However, when $K=20$, we observe significant performance degradation in most LLM invocations, a trend also observed for DTA. 
    This is surprising because increasing the number of indexes should in general reduce workload execution time.
    Moreover, we observe a large performance variation with configurations recommended by different LLM invocations.
    \item\textbf{Real-M}: For all constraints $K\in\{5, 10, 20\}$ tested, LLM-driven index tuning leads to significantly longer workload execution times compared to DTA. 
    \item\textbf{Real-R}: LLM-driven index tuning exhibits substantial performance variation. Although the best responses outperform DTA, the worst responses lead to much longer execution times. With $K=5$ or $K=20$, the worst LLM outcomes achieve almost no improvement over the original configuration.
    \item\textbf{Real-S}: Of all the workloads evaluated, LLM performs the worst for \textbf{Real-S}. Most LLM responses yield little improvement over the original index configuration, with only a few exceptions. In contrast, DTA achieves substantially better performance.
\end{itemize}

\begin{observation}
Overall, DTA provides more stable and reliable improvements for multi‑query workloads while LLM’s performance varies considerably across workloads and invocations. Nevertheless, there are cases in which LLM can significantly outperform DTA.
\end{observation}

\revision{This observation is based on workloads constructed using the top 10 queries with the largest single-query improvements using LLM for \textbf{Real-D} and \textbf{Real-M}. Therefore, including other less beneficial queries will further reduce the relative effectiveness of LLM.} 
We next present two case studies to examine the factors behind LLM's successes and failures on multi-query workloads.


\vspace{-0.5em}
\subsection{Case Study: What Went Right for Real-D?}
\label{sec:multi-query-workloads:real-d}

We first take a closer look at \textbf{Real-D} with $K=10$. 
In this setting, GPT-5 delivers exceptionally strong performance, where all five invocations substantially outperform DTA, with the best achieving $4\times$ further speedup. For the remainder of our analysis, we use this best-performing trial as a representative result for LLM.

\vspace{-0.5em}
\paragraph*{Is the Comparison Fair?}
Because we constrain the LLM by the number of indexes rather than by storage space, a natural question is whether its performance gains arise from generating disproportionately large indexes. To examine this, we measured the storage footprint of the index recommendations produced by both systems. We found that DTA’s indexes occupy 13.7 GB of storage, while LLM's indexes occupy only 7.7 GB. This shows that LLM does not achieve its performance by inflating the storage, indicating that its recommendation is indeed superior in this case.

\begin{figure}[t]
\centering
\subfigure[Execution time of all queries.]{ \label{fig:multi-query:real-d:all}
    \includegraphics[width=.64\columnwidth]{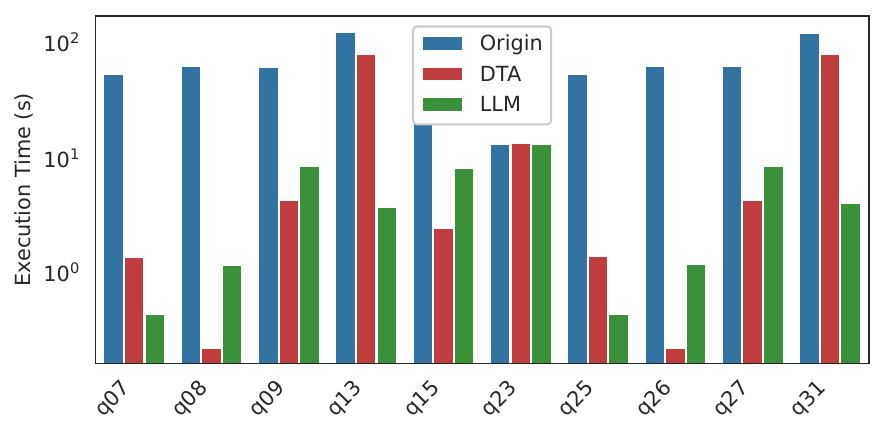}}
\subfigure[Variations of DTA.]{ \label{fig:multi-query:real-d:dtallm}
    \includegraphics[width=0.315\columnwidth]{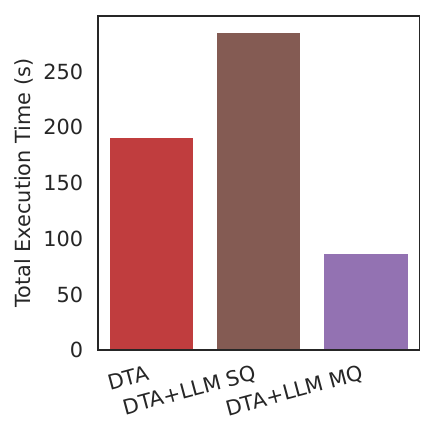}}
\vspace{-1.5em}
\caption{Analysis of \textbf{Real-D} $K=10$.} 
\label{fig:multi-query:real-d:analysis}
\end{figure}

\vspace{-0.5em}
\paragraph*{Per-query Improvement.}
Beyond total execution time, we also evaluated the benefit of the index recommendations at the per-query level. 
Figure~\ref{fig:multi-query:real-d:all} reports the execution time of each query before and after tuning by both DTA and LLM. We observe that nine out of the ten queries exhibit substantial speedups under both methods. Among these nine queries, LLM outperforms DTA for four queries, while DTA performs better for the remaining five queries. This indicates that both approaches optimize the workload effectively, with each favoring different subsets of queries.

\vspace{-0.5em}
\paragraph*{Cross-query Recommendation.}
To better understand GPT-5's behavior in this case, we examine its underlying reasoning process. Rather than selecting the most promising indexes for each query in isolation, GPT-5 appears to prioritize indexes that can benefit multiple queries. The following excerpt illustrates this behavior (all table and column names are anonymized). We observe similar reasoning patterns across all multi-query workloads.

\par\medskip
\noindent
\makebox[\columnwidth][c]{%
  \colorbox[gray]{0.9}{%
    \parbox{\dimexpr\columnwidth-2\fboxsep\relax}{%
      \small\textit
      {... Indexing Tab$_1$ on (Col$_2$, Col$_3$, Col$_4$, Col$_1$) including Col$_5$ could \textcolor{red}{benefit multiple queries, especially $Q_1$, $Q_4$ and $Q_5$} ... I need to limit the proposals to 10 indexes, considering one that \textcolor{red}{aids both queries} on Tab$_2$ ...}
    }%
  }%
}
\par\medskip

To assess the effectiveness of this intention, we analyze the contribution of each index in LLM’s recommendation. Specifically, for every index, we count how many queries actually use it. We find that the top three most frequently used indexes appear in 9, 7, and 5 of the 10 queries, respectively. 
For comparison, the top three indexes recommended by DTA are used by 5, 4, and 3 queries. 

\vspace{-0.5em}
\paragraph*{Improve DTA with LLM}
Furthermore, we find that none of the three most beneficial LLM-recommended indexes is generated as candidates by DTA. 
This observation raises an interesting question: can LLM-recommended indexes enrich the candidate pool of DTA and thus improve its tuning performance? 
To investigate this, we modify DTA’s candidate index generation process to incorporate LLM’s recommendations. 
Figure~\ref{fig:multi-query:real-d:dtallm} presents the results, where ``DTA+LLM SQ'' and ``DTA+LLM MQ'' are two variants that incorporate LLM’s recommendations for each individual query (as in Section~\ref{sec:single-query-workloads}) and for the entire workload, respectively. The result shows that LLM’s multi-query workload recommendations improve DTA’s performance by more than $2\times$. However, including only LLM’s single-query recommendations fails to improve performance and, in fact, significantly slows down overall execution time. 
This degradation again results from inaccurate cost estimates. 

\begin{observation}
Given a multi‑query workload, LLM tends to recommend indexes that can benefit multiple queries, which may produce high‑quality candidate indexes that DTA misses.
\end{observation}

This finding points to an opportunity to improve DTA by expanding its candidate pool with LLM-recommended indexes. However, as we show in Section~\ref{sec:lesson:integration}, directly incorporating such recommendations often leads to performance degradation. Nevertheless, this highlights the potential for integrating LLM's insights into DTA if cost estimation can be further improved, an area that continues to receive active research attention.

\begin{figure}[t]
\centering
\subfigure[Variations of LLM.]{ \label{fig:multi-query:real-m:dtallm}
    \includegraphics[width=0.315\columnwidth]{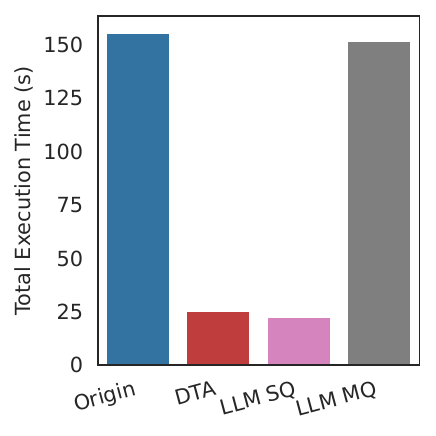}}
\subfigure[Impact of workload size to dominant query tuning.]{ \label{fig:multi-query:real-m:workload-size}
    \includegraphics[width=.64\columnwidth]{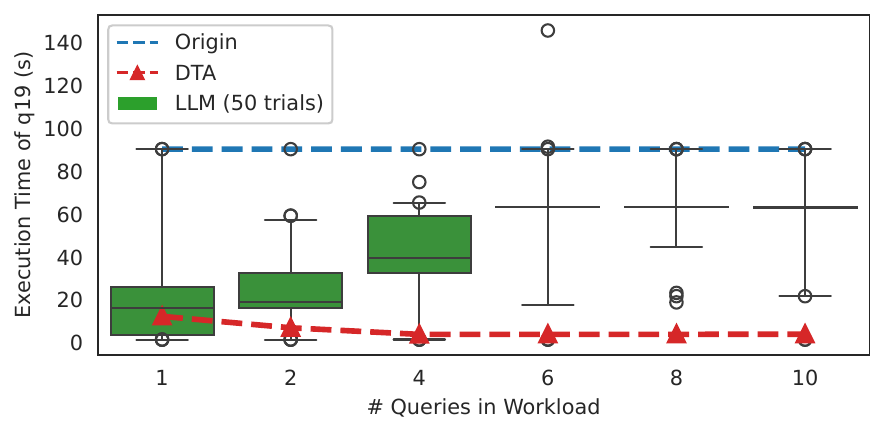}}
\vspace{-1.5em}
\caption{Analysis of \textbf{Real-M} $K=10$.} 
\label{fig:multi-query:real-m:analysis}
\end{figure}

\vspace{-0.5em}
\subsection{Case Study: What Went Wrong for Real-M?}
\label{sec:multi-query-workloads:real-m}

Next, we study the case of \textbf{Real-M} with $K=10$. As shown in Figure~\ref{fig:multi-query:real-m}, all five LLM invocations consistently fail to achieve improvements comparable to those of DTA.

\vspace{-0.5em}
\paragraph*{Recommendation Analysis.}
Unlike \textbf{Real-D} where all queries (without tuning) have relatively similar execution time (Figure~\ref{fig:multi-query:real-d:all}), \textbf{Real-M} is dominated by two queries (i.e., queries 19 and 22), together accounting for roughly 85\% of the total execution time (60\% and 25\%, respectively). Consequently, effective workload optimization hinges on whether LLM can identify these bottleneck queries and recommend indexes to improve them. The five invocations of GPT-5 clearly fail in this regard. 
Among the five responses, only one proposes indexes that can be used by query 19, and only two have impact on query 22. 
Moreover, these responses fail to target the real bottlenecks. In contrast, nearly 90\% of DTA’s improvement for the entire workload comes from these two queries.

We further compare the quality of LLM’s multi-query recommendations with those produced when each query is provided in isolation (as in Section~\ref{sec:single-query-workloads}).
Specifically, we collect all the indexes recommended by LLM from five multi-query responses and, separately, five single-query responses, forming two pools of candidates. 
To obtain a final recommendation within the constraint, for each pool of candidates, we use DTA’s configuration enumeration algorithm (excluding DTA’s own candidate indexes) to select the best 10 indexes. Figure~\ref{fig:multi-query:real-m:dtallm} reports the results, where configurations derived from the multi-query and single-query settings are denoted as ``LLM MQ'' and ``LLM SQ,'' respectively.
We can see that while LLM MQ barely improves workload performance, LLM SQ achieves more than 80\% improvement, reaching a level comparable to DTA.

\vspace{-0.5em}
\paragraph*{Explaining LLM's Poor Performance}
\cut{The above observation points to the possibility that LLM's poor performance in this case may be due to a form of distraction. That is, instead of focusing on the most costly queries, it attempts to identify patterns that improve the workload as a whole and, as a result, overlooks the actual bottlenecks.}
The above observation suggests that instead of focusing on the most costly queries, the LLM attempts to identify patterns that improve the workload as a whole and, as a result, overlooks the actual bottlenecks.

To examine this hypothesis, we vary the workload size and measure its impact on the most costly query (i.e., query 19). Starting from a single-query workload containing only query 19, we incrementally add additional queries to create workloads of sizes \{2, 4, 6, 8, 10\}. 
For each workload, we collect 50 LLM responses to analyze the distribution of their impacts on query 19. 
Figure~\ref{fig:multi-query:real-m:workload-size} summarizes the results, with whiskers indicating the 5th and 95th percentiles. 
For reference, the figure also includes the execution time of query 19 without tuning and with DTA tuning, denoted by the blue and red dashed lines, respectively.

When the workload contains only query 19, roughly 50\% of LLM invocations achieve performance comparable to or better than DTA. 
This fraction drops sharply by adding only one more query, and falls below 5\% once the workload includes four queries. 
The median improvement of query 19 increases monotonically as more queries are added, stabilizing at a workload size of six.  
This result indicates that LLM quality degrades when more queries are included, despite the fact that the estimated costs in the prompt encode the relative importance of each query.  
In contrast, under the cost-based tuning framework, DTA’s improvement for query 19 remains stable and even improves slightly as the workload grows, perhaps due to additional statistics created while tuning other queries.

\begin{observation}
\revision{When information of multiple queries are provided in a plain, concatenated form without additional selection or processing, LLM places less emphasis on the most costly queries leading to suboptimal index recommendations.}
\end{observation}

Note that the above behavior is not unique to index tuning or to the GPT-5 model. Previous work reports similar tendencies in various LLMs when exposed to irrelevant contexts~\cite{pmlr-v202-shi23a}, or long contexts that are relevant but overly extensive~\cite{du-etal-2025-context}, across a range of NLP tasks. Our result shows that such degradation in LLM quality with a larger number of queries can become a serious issue when the underlying workload is complex and unbalanced. 
\revision{Existing work on workload compression~\cite{SiddiquiJ00NC22,Wred} and cost-based prompt construction~\cite{GiannakourisT25,LLMIdxAdvis} potentially offers alternatives  for mitigating this issue (see Sections~\ref{sec:discussion} and~\ref{sec:related-work}).} 
\vspace{-0.5em}
\section{Lessons from LLM-DTA Integration}\label{sec:lesson}

We show that it is nontrivial to integrate LLM with DTA, with two attempts.
For simplicity, we focus on directly integrating LLM-recommended indexes (Sections~\ref{sec:single-query-workloads} and~\ref{sec:multi-query-workloads}) with DTA. Integrating LLM-derived insights (Section~\ref{sec:llm-efficacy}) leads to similar challenges.

\begin{table}[t]
\caption{Performance impact on DTA from including LLM-recommended indexes as additional candidates.}
\vspace{-1em}
\label{tab:llm-dta-integrate}
\small
\centering
\begin{tabular}{|l|r|r|r|r|}
\hline
& \multicolumn{3}{c|}{Single-query (\# queries)} & Multi-query \\
\hline
Workload & Improved & Unchanged & Degraded & Improvement \\
\hline
\hline
TPC-H	&	6	&	12	&    4   & 15.37\%		\\
Real-D	&	10	&	8	&    11  & 54.64\%		\\
Real-M 	&	8	&	18	&    14  & 1.29\%		\\
Real-R 	&	6	&	9	&    9   & -87.33\%		\\
Real-S	&	0	&	6	&    6   & -71.06\%	    \\
\hline
\end{tabular}
\end{table}

\vspace{-0.5em}
\subsection{Cost-based Integration}\label{sec:lesson:integration}

One plausible way to leverage LLM's capability is to integrate its recommendations into the existing cost-based framework of DTA. 
An implementation of this approach has been demonstrated in Section~\ref{sec:multi-query-workloads:real-d} with the \textbf{Real-D} workload, using LLM to only enrich DTA's pool of candidate indexes while keeping the other components of DTA unaffected. 
This integration preserves DTA’s contract to select the configuration with the lowest estimated cost, while expanding its search space with LLM-recommended indexes. 

Table~\ref{tab:llm-dta-integrate} summarizes the evaluation results of this implementation for all workloads, comparing the final configurations found by DTA with and without the LLM-recommended indexes. 
For single-query workloads, we report the number of queries whose execution time improves by at least $5\%$ (``Improved''), remains within $\pm 5\%$ (``Unchanged''), or degrades by at least $5\%$ (``Degraded''). 
For multi-query workloads, we report the overall performance change measured by the total execution time. We have the following finding.

\begin{observation}
Enriching DTA’s pool of candidates with LLM‑recommended indexes does not consistently improve performance, and often leads to performance degradation.
\end{observation}

This frequent performance degradation can again be attributed to inaccurate cost estimates. Despite decades of progress made, cost estimation remains a challenging problem with active research. Advances in this direction may help mitigate the issue if adopted in production. 
Here, we emphasize the practical impact of cost estimation errors (stemming from inaccurate cardinality estimates) in the context of integrating LLM with DTA. 
As a production-grade tuner, DTA serves as a substantially strong starting point and is therefore more sensitive to inaccuracies in cost estimation.

\begin{figure}[t]
\centering
    \includegraphics[width=0.95\columnwidth]{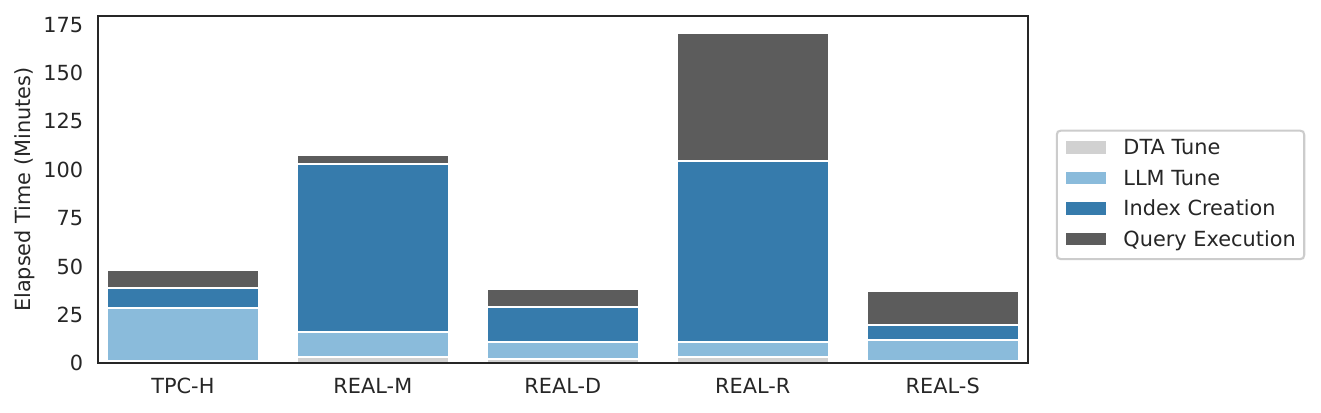}
\vspace{-1em}
\caption{Time breakdown for performance validation.} 
\label{fig:time-breakdown}
\end{figure}

\vspace{-0.5em}
\subsection{Validation-based Integration}\label{sec:lesson:cost}

Another approach is to incorporate performance validation. 
One can materialize the recommended configurations (from both DTA and multiple LLM invocations), execute the corresponding queries, and select the configuration with the best execution time.

Figure~\ref{fig:time-breakdown} shows the end-to-end time breakdown of this approach across all multi-query workloads. Specifically, ``DTA Tune'' and ``LLM Tune'' report the tuning time of DTA and the total time of the five GPT-5 invocations, respectively. ``Index Creation'' and ``Query Execution'' represent the combined index-building and workload-execution time for the configurations recommended by DTA and by the five LLM runs. Different configurations may contain overlapping indexes, and we build each distinct index only once. We first evaluate the recommendation by DTA, followed by the five LLM-generated configurations, using the minimum observed execution time as the timeout for subsequent validations. 

\begin{observation}
The cost of performance validation of recommended configurations is often significantly higher than the cost of index tuning itself, largely due to the overhead of index creation.
\end{observation}

Previous work has identified similar challenges and proposed mitigation techniques that focus primarily on reducing query execution cost~\cite{GiannakourisT25,lim2024hit} or improving data efficiency~\cite{Wu25}. Here, we highlight the overhead of index creation, which can become a more substantial bottleneck and is easily amplified by the diverse recommendations that LLM could produce.

\section{Future Opportunities}\label{sec:discussion}


\paragraph*{Enhancing LLM-driven Index Tuning} 
In this paper, we evaluate pre-trained LLM for index tuning in an end-to-end setting, using only basic database and workload information. 
Beyond this setup, there are several opportunities to further leverage LLM for index tuning. 
\revision{For example, prompts can be enriched with additional information, such as DTA recommended indexes for the same query or similar queries, which can serve as reference demonstrations~\cite{LLMIdxAdvis}. 
In addition, LLM can be adapted to the index tuning task through task-specific fine-tuning~\cite{sft,rft-dpo,rft-ppo,rft-grpo}. 
Moreover, the tuning process can be decomposed into multiple tasks and handled by a multi-agent framework~\cite{multi-agent-JhaCLDKJ25,multi-agent-LiXLTDL25,MAAdvisor}, where existing index tuners such as DTA can be integrated as an external tool to provide additional guidance and/or feedback at different stages.}
\revision{Understanding how these enhancements impact real-world workloads, particularly in addressing the limitations of LLM observed in our study, such as large performance variance and the tendency toward distraction, remains an interesting direction for future work.}

\vspace{-0.5em}
\paragraph*{Distilling Insights from LLM} 
As demonstrated in Section~\ref{sec:llm-efficacy}, heuristics used by LLM could be distilled and exploited to narrow the gap between existing index tuners and the best of LLM. 
Whether additional insights can be extracted and applied more systematically remains an open question. 
Beyond integrating these insights into existing index tuners, an alternative is to distill the knowledge into smaller, task-specific models~\cite{gu2024minillm,mansourian2025a,yang2025survey}. Such models may provide a more efficient and predictable option for production deployment while retaining much of the effectiveness of LLM. 

\vspace{-0.5em}
\paragraph*{Making Performance Validation a First-class Citizen in Index Tuning} 
A key roadblock in using LLM-recommended indexes with today’s cost-based index tuners is the inaccuracy of cost estimation. 
As a result, incorporating performance validation as a first-class citizen appears increasingly important to production index tuners. 
Some recent industrial systems~\cite{DasGIJJNRSXC19,YadavVZ23,ChakkappenKMKSZZLZ25} have incorporated validation mechanisms, but typically only to verify the final configuration selected by the tuner. This restriction is largely due to the substantial cost of validation. 
As shown in Section~\ref{sec:lesson:cost}, incorporating LLM-driven index tuning further amplifies this challenge, as LLM can easily generate a large and diverse set of candidate recommendations, significantly increasing the need for validation. Therefore, an important direction for future work is to develop low-overhead while less disruptive~\cite{neon-db-branching} validation techniques. 
A further step is to integrate this validation component more tightly into the index tuning process, allowing validation outcomes to complement existing cost estimates and directly influence tuning decisions. 
\cut{This would require reformulating the index tuning problem to explicitly model the validation cost as an additional budget constraint, offering a practical way to balance tuning quality against validation overhead.}
\section{Related Work}
\label{sec:related-work}

\paragraph*{LLM-driven Index Tuning} A flurry of recent work has been proposed to unlock the power of LLMs to improve the performance of database systems~\cite{LaoWLWZCCTW25,Trummer24,GiannakourisT25}.
Most of this line of work is devoted to tuning configuration parameters (a.k.a. knobs) of database systems, which itself is a fast-growing research area with intensive studies~\cite{DuanTB09,AkenPGZ17,CeredaVCD21,KanellisDKMCV22,LiZLG19,ZhangLZLXCXWCLR19,AkenYBFZBP21,WangTB21}.
To the best of our knowledge, so far there has been little work on using LLM for index tuning yet, except for a couple of very recent proposals, such as $\lambda$-Tune~\cite{GiannakourisT25}, LLMIdxAdvis~\cite{LLMIdxAdvis}, and MAAdvisor~\cite{MAAdvisor}.
Both $\lambda$-Tune and LLMIdxAdvis follow the end-to-end workflow evaluated in our study, while MAAdvisor proposes a different multi-agent framework with a much more complicated workflow.
Unlike the classic cost-based index tuning that has been well studied and evaluated in real-world industrial deployments~\cite{DasGIJJNRSXC19,YadavVZ23,ChakkappenKMKSZZLZ25}, LLM-driven index tuning remains an underexplored area, especially with an industrial setup.

\vspace{-0.5em}
\paragraph*{Workload Compression}
\revision{
As we demonstrated in Section~\ref{sec:multi-query-workloads}, LLM-driven index tuning does not have an advantage over DTA when tuning multi-query workloads and workload size does have an impact on the performance of LLM.
Therefore, workload compression~\cite{ChaudhuriGN02,DeepGKNV20,SiddiquiWNC22,Wred} will continue to be an interesting and critical direction for future work to improve the efficacy of LLM for tuning multi-query workloads, which has also explored by $\lambda$-Tune~\cite{GiannakourisT25}.
Recent work on foundation database models~\cite{DBLP:conf/cidr/WehrsteinB0VGW25} proposes learning compact representations of datasets and workloads in a latent space that generalize across tasks, 
which could also be used to guide workload compression for LLM-based index tuners.}

\vspace{-0.5em}
\paragraph*{Robustness of LLMs}
In addition to database performance tuning (including index tuning), LLMs have also been applied to other aspects of improving database performance, such as performance diagnosis~\cite{GiannakourisT24,ZhouLSLCWLFZ24} and query optimization~\cite{abs-2411-02862,SunZLYFZ25}.
One common issue found by this line of work is the robustness of LLM invocations, which often respond with incorrect or irrelevant results due to problems such as hallucination.
We have also highlighted such robustness issues for LLM-driven index tuning in Section~\ref{sec:single-query-workloads:robustness}, which calls for future research effort.
One solution to improve robustness, as was proposed by $\lambda$-Tune, is to include a postprocessing validation stage that creates the recommended indexes by LLM and executes the workload queries to identify the best indexes.
Although this solution is guaranteed to find the best indexes among the recommended candidates, it incurs considerable overhead that is perhaps infeasible in practical industrial applications, as shown in Section~\ref{sec:lesson:integration}. 
Therefore, the development of less costly validation technologies is another area of interest for future work.


\section{Conclusion}
\label{sec:conclusion}

We empirically evaluated the effectiveness of LLM-driven index tuning on real-world and industry benchmark workloads on Microsoft SQL Server, and compared it with a commercial index tuner, Database Tuning Advisor (DTA).
 Our results reveal that LLMs demonstrate the potential to provide high-quality index recommendations, particularly in cases where DTA's recommendations are not effective due to errors in the query optimizer's cost estimation. 
 However, LLMs exhibit substantial variance in the quality of index recommendations, both across invocations for the same query and across queries. Furthermore, incorporating LLM-recommended indexes as additional candidates in DTA's search space yielded no significant improvements in quality. Thus, despite their promise, the use of LLMs for index tuning in practice remains an open and challenging problem. 
 Based on our findings, we identify potential areas of future work that aim to make LLM-driven index tuning more practical, such as techniques that can reduce the variance of LLMs without significantly reducing its quality, as well as more efficient mechanisms for validating index recommendations.


\clearpage

\balance
\bibliographystyle{ACM-Reference-Format}
\bibliography{paper}

\clearpage
\appendix
\section{More Evaluation Results}

\subsection{More Results of LLM}

\begin{figure*}
\setcounter{subfigure}{0}
\subfigure[\textbf{TPC-H}]{ \label{fig:llm-vs-dta-extended:tpch}
    \includegraphics[width=0.47\textwidth]{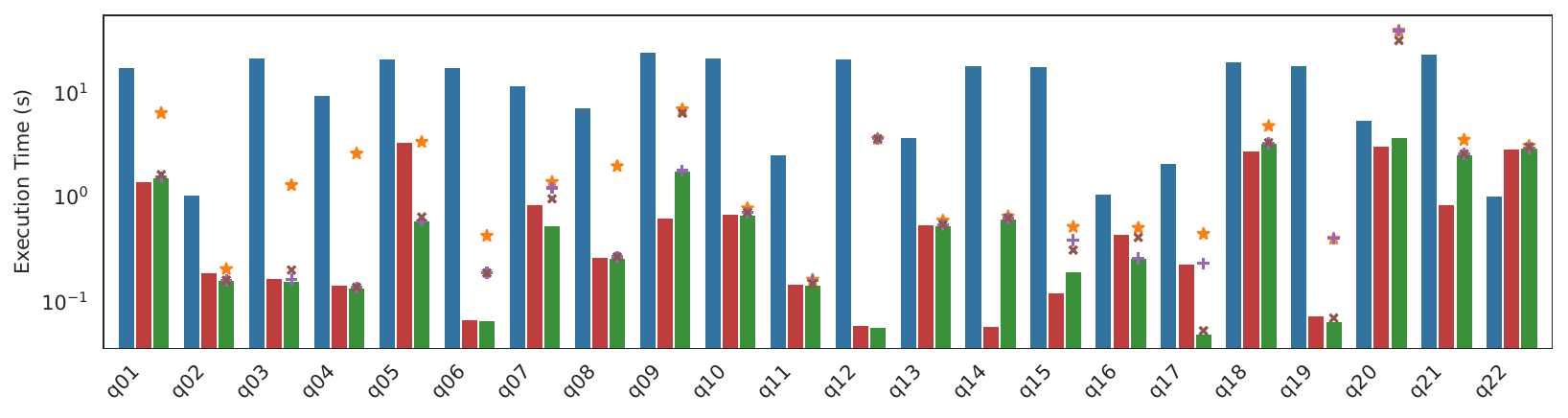}}
\setcounter{subfigure}{3}
\subfigure[\textbf{Real-R}]{ \label{fig:llm-vs-dta-extended:real-r}
    \includegraphics[width=0.51\textwidth]{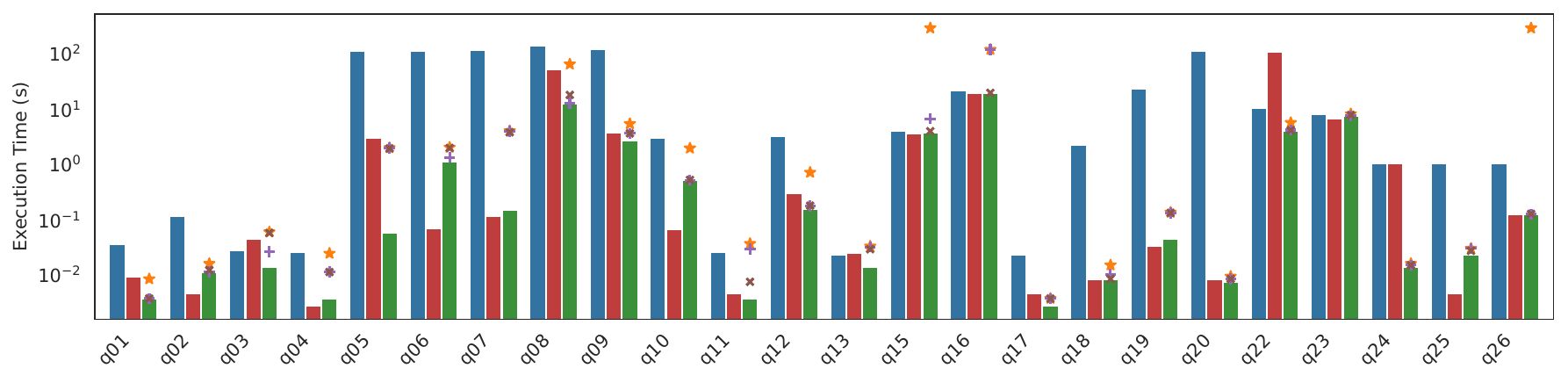}}
\setcounter{subfigure}{1}
\subfigure[\textbf{Real-D}]{ \label{fig:llm-vs-dta-extended:real-d}
    \includegraphics[width=0.69\textwidth]{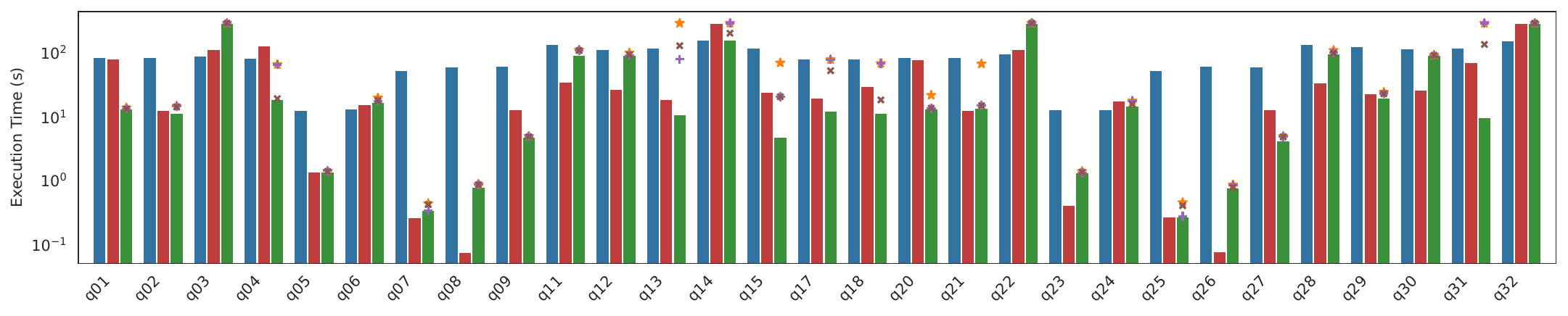}}
\setcounter{subfigure}{4}
\subfigure[\textbf{Real-S}]{ \label{fig:llm-vs-dta-extended:real-s}
    \includegraphics[width=0.28\textwidth]{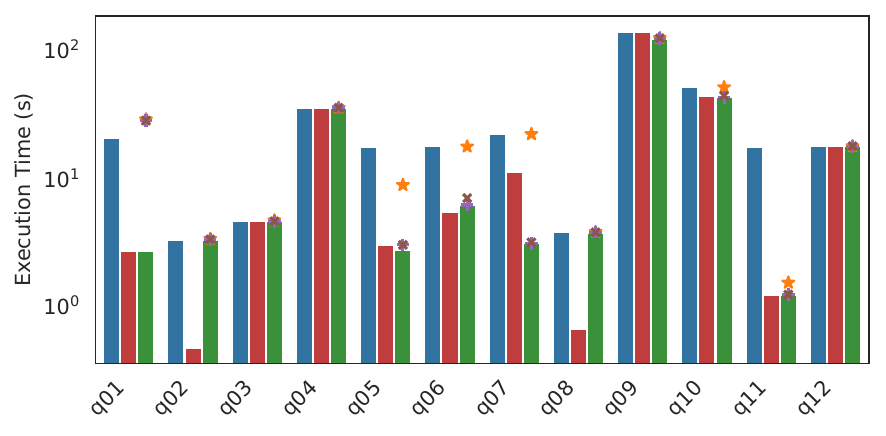}}
\setcounter{subfigure}{2}
\subfigure[\textbf{Real-M}]{ \label{fig:llm-vs-dta-extended:real-m}
    \includegraphics[width=0.98\textwidth]{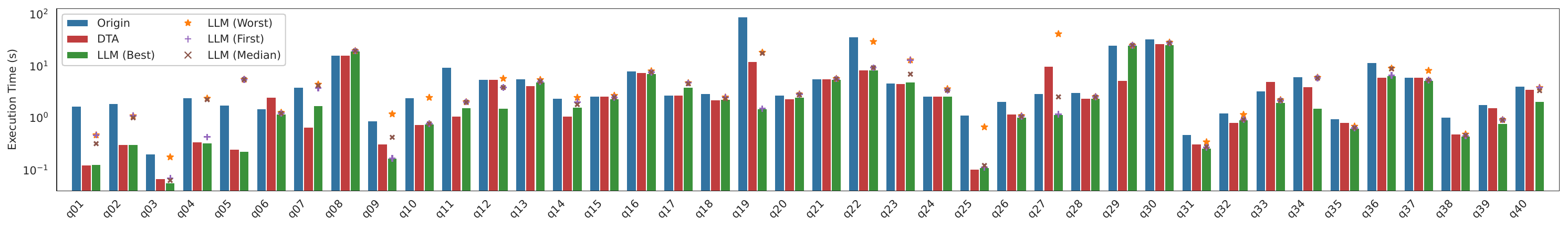}}
\vspace{-1.5em}
\caption{\revision{LLM-driven index tuning vs. DTA for tuning single-query workloads.}}
\label{fig:llm-vs-dta-extended}
\end{figure*}

\revision{In addition to the best and worst results among the five LLM invocations, we also consider the first and median results, denoted by $+$ and $\times$ in Figure~\ref{fig:llm-vs-dta-extended}, respectively. Both settings simulate the scenario where a single LLM response is directly adopted without additional performance validation.}

\vspace{-1em}
\revision{\begin{itemize}[leftmargin=*]
    \item\textbf{TPC-H}: Unlike the worst response, where LLM fails to outperform DTA for any query, the first and median responses still identify better recommendations for some queries, such as queries 5 and 17. However, they both exhibit substantially worse performance than the best response for several other queries (e.g., queries 9, 12, 15, and 20).
    \item\textbf{Real-D}: For queries with large performance variance in five LLM invocations (i.e., where the best response significantly outperforms DTA while the worst significantly underperforms it), the first and median responses generally fail to outperform or only marginally match DTA (e.g., queries 13, 15, 17, 18, and 31).
    \item\textbf{Real-M}: Similar to \textbf{Real-D}, the first and median responses often fail to outperform DTA and do not capture the gains observed in the best-case results, particularly for queries with high variance across different LLM invocations (e.g., queries 1, 2, 5, 7, 12, and 34). Nevertheless, they still identify better recommendations than DTA for a few queries, such as query 19 (first response) and query 27 (both first and median responses).
    \item\textbf{Real-R}: We observe that the two significant QPRs caused by the worst LLM responses (i.e., queries 15 and 26) no longer result in timeouts with the first and median responses. They also achieve performance comparable to, or even better than, DTA for some high-variance queries (e.g., queries 8 and 26). However, they perform significantly worse than DTA for a few queries where the best LLM response previously matched or outperformed DTA (e.g., queries 7 and 11).
    \item\textbf{Real-S}: The first and median LLM responses match the best response in most cases, except for query 1, where they perform significantly worse than DTA and result in a QPR.
\end{itemize}}

\revision{In summary, the above results indicate that directly adopting a single LLM response without further validation often leads to suboptimal performance and generally underperforms DTA. 
Although the first and median responses still markedly outperform DTA (i.e., more than 5\% better) for 26\% and 29\% of the queries, respectively, more than half of the 127 queries exhibit noticeable degradation (i.e., more than 5\% worse) in both settings. In contrast, the best response achieves such degradation for only about 30\% of the queries. 
This contrast highlights the importance of validating LLM-generated recommendations before adoption.}

\subsection{More Results of Rule-based Tuner}

\subsubsection{End-to-end Evaluation}

\begin{figure*}
\setcounter{subfigure}{0}
\subfigure[\textbf{TPC-H}]{ \label{fig:simple-all:tpch}
    \includegraphics[width=0.47\textwidth]{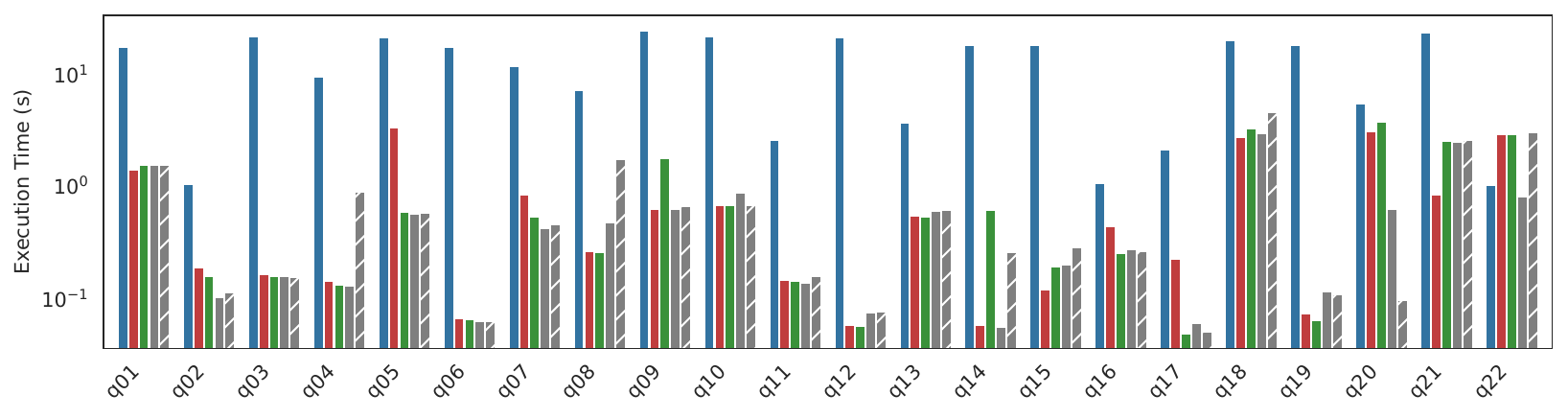}}
\setcounter{subfigure}{3}
\subfigure[\textbf{Real-R}]{ \label{fig:simple-all:real-r}
    \includegraphics[width=0.51\textwidth]{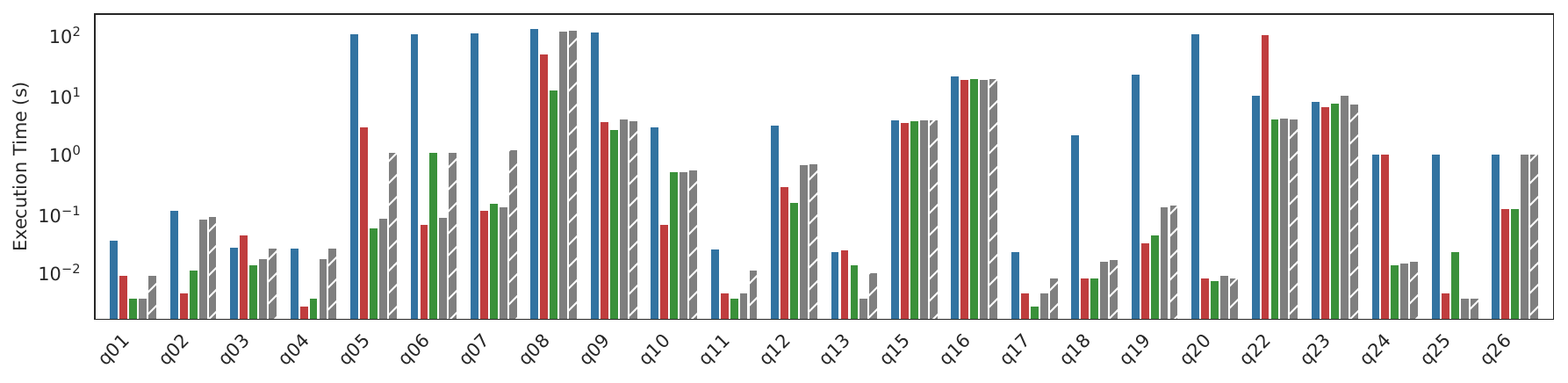}}
\setcounter{subfigure}{1}
\subfigure[\textbf{Real-D}]{ \label{fig:simple-all:real-d}
    \includegraphics[width=0.69\textwidth]{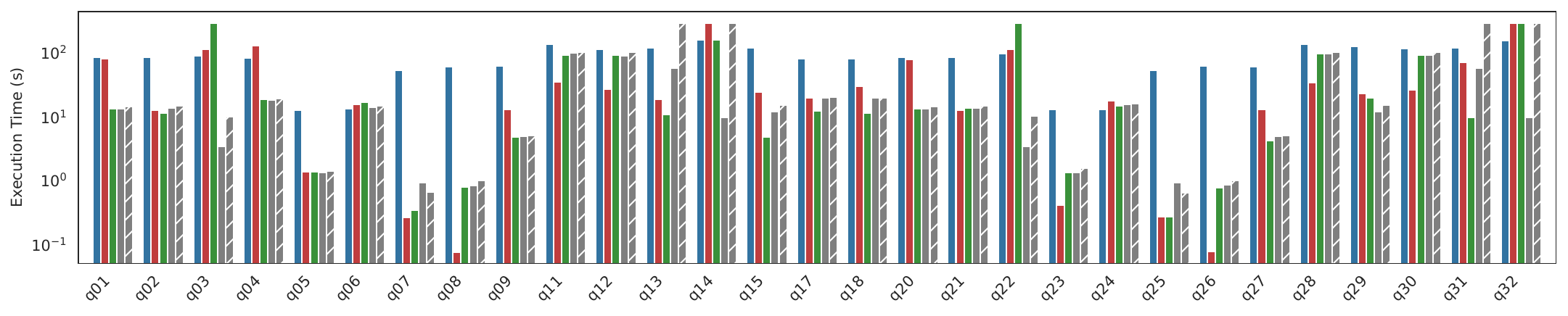}}
\setcounter{subfigure}{4}
\subfigure[\textbf{Real-S}]{ \label{fig:simple-all:real-s}
    \includegraphics[width=0.28\textwidth]{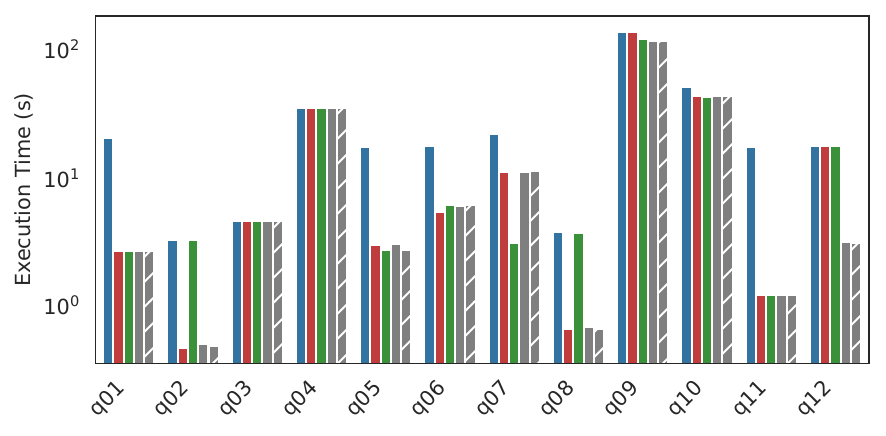}}
\setcounter{subfigure}{2}
\subfigure[\textbf{Real-M}]{ \label{fig:simple-all:real-m}
    \includegraphics[width=0.98\textwidth]{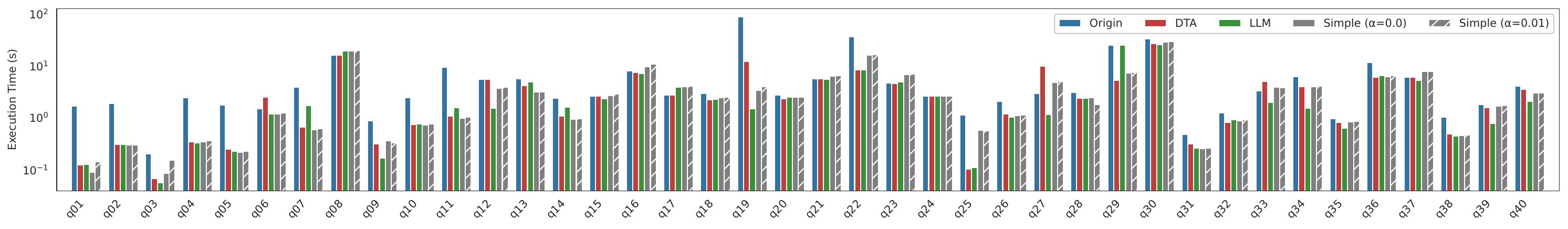}}
\vspace{-1.5em}
\caption{\revision{End-to-end evaluation of simple rule-based index tuner for tuning single-query workloads.}}
\label{fig:simple-all}
\vspace{-1.5em}
\end{figure*}

\revision{In this section, we present the complete evaluation of the rule-based tuner introduced in Section~\ref{sec:llm-efficacy:simple}, comparing it with DTA and LLM (best result in five invocations). Figure~\ref{fig:simple-all} summarizes the results, where the gray and hatched gray bars correspond to the simple rule-based tuner with $\alpha=0$ and $\alpha=0.01$, respectively.}

\vspace{-1em}
\revision{\begin{itemize}[leftmargin=*]
    \item\textbf{TPC-H}: With $\alpha=0$, the simple rule-based tuner achieves performance comparable to, and in some cases better than, DTA, with only a few exceptions (e.g., queries 8, 12, 15, 19, and 21). It also underperforms LLM for only five queries (e.g., queries 8, 10, 12, 17, and 19). Notably, it outperforms both DTA and LLM for query 20 by approximately $5\times$.
    When $\alpha$ increases to $0.01$, performance degrades on several queries, including queries 4, 8, 14, and 22. In particular, for query 4, the tuner fails to match the performance of DTA and LLM, becoming about $6\times$ slower. On the other hand, performance on query 20 further improves, achieving around $30\times$ speedup over both DTA and LLM.
    \item\textbf{Real-D}: There are four queries (i.e., queries 3, 14, 22, and 32) where neither DTA nor LLM identifies effective recommendations to improve performance, whereas the rule-based tuner ($\alpha=0$) discovers configurations that yield improvements by up to $33\times$. Meanwhile, it significantly underperforms both DTA and the LLM for some queries (e.g., queries 7, 13 and 25). Furthermore, with $\alpha=0.01$, the rule-based tuner times out for four queries (i.e., queries 13, 14, 31, and 32).
    \item\textbf{Real-M}: Unlike on \textbf{Real-D}, the rule-based tuner does not yield substantial improvements over DTA and LLM. In fact, it underperforms both baselines for a few queries (e.g., queries 16, 22, 25, and 37), while achieving comparable performance in most other cases. 
    The performance of the rule-based tuner is not sensitive to the choice of threshold $\alpha$ in this workload.
    \item\textbf{Real-R}: The observation is similar to that for \textbf{Real-M}: the rule-based tuner with $\alpha=0$ achieves performance comparable to both DTA and LLM for more than half of the queries. However, it underperforms both baselines for several queries (e.g., queries 2, 4, 8, 12, 18, 19, and 26). Increasing $\alpha$ to $0.01$ further degrades performance for three queries (i.e., queries 5, 6, and 7).
    \item\textbf{Real-S}: The simple rule-based tuner (with both $\alpha=0$ and $0.01$) significantly outperforms both DTA and LLM for query 12. It matches DTA for the rest of the queries and clearly outperforms LLM for queries 2 and 8.
\end{itemize}}

\revision{Overall, despite its simplicity, the rule-based tuner (with $\alpha=0$) is surprisingly effective as a standalone approach for single-query workloads. It achieves performance comparable to both DTA and LLM (best of five invocations) in most cases and occasionally identifies configurations that outperform both baselines by up to more than $30\times$. However, it is not sufficient to replace either DTA or LLM, as it underperforms both on a substantial fraction of queries (i.e., 37\% and 41\% for DTA and LLM, respectively).
Specifically, across all the 127 queries, it outperforms, underperforms and falls within the range of the two baselines (within $\pm 5\%$) for 16\%, 23\% and 61\% of queries, respectively. 
With $\alpha=0.01$, performance remains similar for most queries, but degrades substantially for some, in a few cases leading to timeouts.}

\subsubsection{Recommendation Cross-check}

\begin{table}[t]
\caption{\revision{Overlap between LLM (five invocations) and rule-based index recommendations.}}
\vspace{-1em}
\label{tab:simple-coverage}
\small
\centering
\begin{tabular}{|l|r|r|r|r|r|}
\hline
& & \multicolumn{2}{c|}{$\alpha=0$} & \multicolumn{2}{c|}{$\alpha=0.01$} \\
\hline
Workload & \# LLM & \# Total & \#(\%) Matched & \# Total & \#(\%) Matched \\
\hline
\hline
TPC-H	&	200	&	72	&    36 (50\%)   & 47 & 32 (68\%)		\\
Real-D	&	937	&	374	&    106 (28\%)  & 104 & 47 (45\%)	\\
Real-M 	&	507	&	640	&    100 (16\%)  & 156 & 81	(52\%)	\\
Real-R 	&	365	&	100	&    57 (57\%)  & 32 & 21 (66\%)		\\
Real-S	&	93	&	116	&    17 (15\%)   & 13 & 8 (62\%)	    \\
\hline
\end{tabular}
\end{table}

\revision{We further conduct a cross-check between indexes recommended by the rule-based tuner and those recommended by the LLM. The results for each workload are summarized in Table~\ref{tab:simple-coverage}. Specifically, “\# LLM” and “\# Total” denote the number of distinct indexes recommended by GPT-5 (across all five invocations) and by the simple rule-based tuner, respectively. “\# Matched” indicates the number of indexes generated by the rule-based tuner that match those recommended by the LLM.
An index from the rule-based tuner is considered a match if its key prefix exactly matches that of an LLM-recommended index and it covers all columns included in the LLM index (i.e., both key columns and included columns). This definition ensures that a matched index from the rule-based tuner can provide at least the same functionality as the corresponding LLM-recommended index. We also report the percentage of matched indexes among those generated by the rule-based tuner in parentheses.}

\revision{We observe that the percentage of matched indexes increases significantly as $\alpha$ becomes larger. This trend suggests that GPT-5 indeed tends to recommend fewer indexes for smaller tables, which are filtered out by the rule-based tuner as $\alpha$ increases. For the remaining larger tables (with $\alpha=0.01$), more than half of the indexes generated by the rule-based tuner are also recommended by the LLM for most workloads, indicating substantial overlap in recommendations. 
At the same time, a noticeable gap remains between the total number of distinct indexes recommended by the LLM and the matched indexes, suggesting that many LLM-recommended indexes are not yet captured by the current set of rules and could potentially be further distilled into additional patterns.}

\end{document}